# Autoigniton of *n*-Butanol at Low to Intermediate Temperature and Elevated Pressure

Bryan William Weber

B.S., Case Western Reserve University, 2009

A Thesis

Submitted in Partial Fulfillment of the

Requirements for the Degree of

Master of Science

at the

University of Connecticut

2010

APPROVAL PAGE

Master of Science Thesis

Autoigniton of *n*-Butanol at Low to Intermediate Temperature and Elevated Pressure

Presented by

Bryan William Weber, B.S.

Major Advisor \_\_\_\_\_\_\_\_\_\_\_\_\_\_\_\_\_\_\_\_\_\_\_\_\_\_\_\_\_\_\_\_\_\_\_\_\_\_\_\_\_\_\_

Chih-Jen Sung

Associate Advisor \_\_\_\_\_\_\_\_\_\_\_\_\_\_\_\_\_\_\_\_\_\_\_\_\_\_\_\_\_\_\_\_\_\_\_\_\_\_\_\_\_\_\_

Baki Cetegen

Associate Advisor \_\_\_\_\_\_\_\_\_\_\_\_\_\_\_\_\_\_\_\_\_\_\_\_\_\_\_\_\_\_\_\_\_\_\_\_\_\_\_\_\_\_\_

Tianfeng Lu

Associate Advisor \_\_\_\_\_\_\_\_\_\_\_\_\_\_\_\_\_\_\_\_\_\_\_\_\_\_\_\_\_\_\_\_\_\_\_\_\_\_\_\_\_\_\_

Michael Renfro

University of Connecticut

2010

Dedicated to my Opa, without whom I would not be the student I am today;

and to my family and friends, without whom I would not be the person I am today

# Acknowledgements

The list of people who deserve thanks for all their help with this project is too long to print here. However, the following people made extraordinary contributions to this work and deserve special recognition.

Dr. Chih-Jen Sung, my advisor, who offered patient and caring guidance through this whole process. Without his help and insistence on "doing it right," this project could not have been completed;

Dr. Kamal Kumar, who helped in more ways than can be counted, but especially with the data collection and analysis;

Dr. Yu Zhang, who provided indispensible help with the GCMS and testing the mixture compositions;

Dr. Bill Green, Professor of Chemical Engineering at Massachusetts Institute of Technology, who graciously provided his group's *n*-butanol reaction mechanism prior to publication;

Dr. Anne Weber, my wonderful mother, who helped revise and edit this document;

Dr. Baki Cetegen, Dr. Tianfeng Lu, and Dr. Michael Renfro, my advisory committee;

Kyle Brady, Apurba Das, Matthew Deans, Billal Hossain, Xin Hui, and Kyle Niemeyer, my lab mates and friends, whose technical and personal support went above and beyond any reasonable expectations;

and all my family, friends, and colleagues, whose encouragement and assistance made this process much easier than it would have been otherwise.

This material is based upon work supported as part of the Combustion Energy Frontier Research Center, an Energy Frontier Research Center funded by the U.S. Department of Energy, Office of Science, Office of Basic Energy Sciences under Award Number DE-SC0001198. The author was also supported by the Graduate Assistantship in Areas of National Need Pre-Doctoral Fellowship.

# Table of Contents

# Autoigniton of *n*-Butanol at Low to Intermediate Temperature and Elevated Pressure

Abstract

by

Bryan William Weber

Autoignition delay experiments were performed for *n*-butanol in a heated rapid compression machine. Experiments were performed at pressures of 15 and 30 bar, in the temperature range 650-900 K, and for equivalence ratios of 0.5, 1.0, and 2.0. Additionally, the initial fuel mole fraction and initial oxygen mole fraction were varied independently to determine the influence of each on ignition delay. Over the conditions studied, it was found that the reactivity of the mixture increased as equivalence ratio, initial fuel mole fraction or initial oxygen mole fraction increased. A non-linear correlation to the experimental data was performed and showed nearly second order dependence on the initial oxygen mole fraction and nearly first order dependence on initial fuel mole fraction and compressed pressure. This was the first study of the ignition of *n*-butanol in this temperature and pressure range, and contributes to a better understanding of the chemistry of this fuel under the conditions relevant to practical devices.

Experimentally measured ignition delays were compared against the ignition delay computed from several reaction mechanisms in the literature. The agreement between experiments and simulations was found to be unsatisfactory. Sensitivity analysis was performed and indicated that the uncertainties of the rate constants of parent fuel decomposition reactions play a major role in causing the poor agreement. Further path analysis of the fuel decomposition reactions supported this conclusion and highlighted the particular importance of certain pathways. Based on these results, it was concluded that further investigation of the fuel decomposition, including speciation measurements, will be required.

# Chapter 1. Introduction

Recent concerns over energy security and the environment have created a renewed push to reduce our dependence on fossil fuels. Since petroleum supplies 95% of the energy used in the transportation sector[1], several initiatives are being explored to replace petroleum with alternative fuels. Biofuels are rapidly moving to the forefront as the most promising option. As defined here, biofuels are those that are produced from biomass. This is in contrast to fossil fuels, which are mined or drilled from the Earth and were produced over millions of years by geological processes. The advantages of biofuels include reduced dependence on foreign sources of energy, reduced greenhouse gas (GHG) emissions, and sustainable production.

## 1.1     Biofuels

The biomass sources for biofuels have traditionally been glucose-rich plants, such as sugar cane, sugar beet and, in the US, corn. Glucose is relatively easy to convert to alcohols (e.g., ethanol) by fermentation with yeast. Unfortunately, less than half of the sugar content of even glucose-rich plants is in the form of a starch that can be converted to glucose[2]. Most of the dry mass of plants is composed of more complex sugars, collectively known as lignocellulose. Lignocellulose includes cellulose, hemicellulose and lignin and primarily makes up the cell walls of plants[3].

Although biofuels are a potential alternative to fossil fuels, current production methods have several limitations. Chief among these are concerns about the use of arable land and competition between fuel production and food sources; as more farmland is needed to produce crops for fuel, less is available to use for food crops[4]. This may drive up the cost of food, especially in the developing world. In addition, current farming techniques require large quantities of fertilizer and pesticides, both of which have strong negative environmental impacts.

To help ameliorate these concerns, second-generation biofuels are being developed that use lignocellulose instead of starch as the primary process feedstock[5].

Switching to production of lignocellulosic biofuel offers the opportunity to use entirely different raw materials than those of current biofuels. These include perennial, fast-growing plants such as switchgrass as well as waste products such as paper or wood chips[2,5]. Most crops used in lignocellulosic biofuel require less fertilizer and water and can be grown on marginal land that cannot support food crops. They can sequester atmospheric carbon in the soil and help produce a carbon-neutral cycle to control GHG release. Using these crops and wastes, lignocellulosic biofuels can be produced sustainably without depleting natural resources.

## 1.2 Ethanol vs. Butanol

The most popular biofuel currently in use is ethanol ($C_2H_6O$), which has been used as a fuel for internal combustion engines since Henry Ford designed a flex-fuel car in 1908[6]. By comparison, one of the most promising second-generation biofuels is butanol ($C_4H_{10}O$), with several advantages over current biofuels such as ethanol. Although ethanol is inexpensive to produce, concerns about its production methods have been discussed above. Because it can be produced from lignocellulose, butanol has advantages over ethanol in its production methods.

Although ethanol has a higher octane number, the octane number of butanol still falls within the range of gasoline. (cf. Table 1.1) However, compared with butanol, ethanol has several disadvantages when used with modern engines designed for gasoline. Ethanol is fully miscible in water, while certain of the isomers of butanol are only slightly soluble in water[7]. Due to this, currently existing fuel transportation infrastructure cannot support ethanol. However, the infrastructure in place can carry butanol without any modification.

Ethanol causes material compatibility problems in the fuel lines and seals of the engine, while butanol is much less corrosive[8,9]. Ethanol has a higher vapor pressure than butanol (Table 1.1)[8,10,11]. Higher vapor pressure leads to evaporative emissions which can cause smog and increase the risk of explosion.

| Fuel | Lower Heating Value [MJ/kg] | Volumetric Energy Density [MJ/L] | Research Octane Number (RON) | Motor Octane Number (MON) | Vapor Pressure @ 25°C [torr] |
|---|---|---|---|---|---|
| Gasoline | 42.5 | 32 | 92-98 | 82-88 | |
| Methanol | 19.9 | 16 | 136 | 104 | 127 |
| Ethanol | 28.9 | 20 | 129 | 102 | 59 |
| 1-Butanol | 33.1 | 29 | 96 | 78 | 6 |

**Table 1.1: Properties of Gasoline, Methanol, Ethanol and 1-Butanol. From Ref. [9].**

As shown in Table (1.1), ethanol has a 38% lower volumetric energy density than gasoline, while butanol is only 9% lower[9]. This significantly improves the volumetric fuel economy (i.e., miles per gallon) of butanol/gasoline mixtures over ethanol/gasoline mixtures. In addition, butanol can be mixed in higher proportions with gasoline than ethanol, because the energy content is more closely matched.

Unfortunately, facilities do not currently exist to create sufficient quantities of butanol to enable its widespread use as a gasoline additive or substitute, although the production of such facilities is in progress[2-5,12-36]. Work is underway to modify microorganisms that can innately process biomass into alcohols and require only small genetic changes to increase production of butanol[12,27]. In addition, other scientists are adding metabolic pathways to produce higher alcohols in *E. coli* by genetic engineering[13,14,35]. Work is also being completed on ways to increase the yield of lignocellulose from plants[2]. In short, significant work has been done to support the production of butanol. However, research into the combustion properties of butanol is still in the early stages.

## 1.3 Properties of Butanol

There are four isomers of butanol – *n*-butanol (1-butanol, butan-1-ol, biobutanol), *sec*-butanol (2-butanol), *iso*-butanol and *tert*-butanol. The structure of the isomers of butanol are shown in Figs. (1.1-1.4). *n*-Butanol, *sec*-butanol and *iso*-butanol are liquids at room temperature, while the melting point of pure *tert*-butanol is about 25°C[37]. Despite this, *tert*-butanol has previously been used as an octane enhancer in gasoline[9].

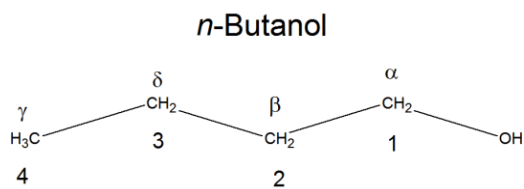

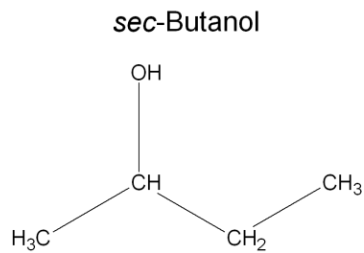

**Figure 1.1: Structure of *n*-butanol**

**Figure 1.2: Structure of *sec*-butanol**

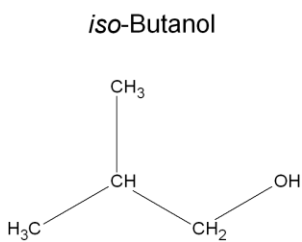

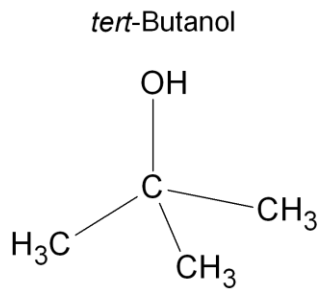

**Figure 1.3: Structure of *iso*-butanol**

**Figure 1.4: Structure of *tert*-butanol**

## 1.4 Butanol Combustion

Of the four isomers of butanol, *n*-butanol has been most extensively studied, although interest is increasing in the other isomers of butanol that have higher octane number than *n*-butanol[28,38]. Several studies have examined *n*-butanol combustion in spark-ignited engines[39-43]. The most recent of those studies was by Zhang et al.[43], who also report species profiles prior to ignition. The species were identified by Gas Chromatography/Mass Spectrometry (GCMS) and quantified by Gas Chromatography-Flame Ionization Detection (GC-FID). Other groups have studied the pyrolysis of various isomers of $C_3$ and $C_4$ alcohols[44-48]. Studies have examined the laminar flame speed and species profiles of *n*-butanol in a jet-stirred reactor for varying equivalence ratios[11,49-51]. Yang et al.[52] studied the four isomers of butanol in low-pressure rich flames and identified some combustion intermediates by vacuum-ultraviolet molecular beam-photoionization-mass spectrometry. Cullis and Warwicker[53] have performed a similar low pressure work using GC to determine slow combustion intermediates of *n*-butanol and report cool flame behavior between 305 and 340°C. Moss et al.[9] presented ignition delays behind reflected shock waves from 1-4 bar and 1200-1800 K for the four isomers of butanol. Black et al.[54] presented shock tube ignition data up to 8 atm and between 1100-1800 K. Several authors have also presented non-premixed flame configurations: Sarathy et al.[11] for *n*-butanol, McEnally and Pfefferle[55] and Grana et al.[56] for the four isomers, and Hamins and Seshadri[57] for $C_1$, $C_2$ and $C_4$ straight-chain alcohols.

Several authors presented theories on the decomposition of the *n*-butanol. McEnally and Pfefferle[55] and Barnard[45] determined that unimolecular decomposition is the primary destruction pathway. Sarathy et al.[11], Moss et al.[9], and Zhang et al.[43] determined that H-atom abstraction is the primary decomposition method. Sarathy et al.[11] postulated that the reason for the difference is the peculiarity of each experimental apparatus, based on the observation that radicals are required for H-abstraction. McEnally and Pfefferle[55] worked on non-premixed coflow flames and Barnard[45] performed pyrolysis. Since there is no oxygen present in pyrolysis, unimolecular decomposition is the only way for the fuel to decompose. McEnally and Pfefferle[55] measure centerline concentration profiles in their flame where the radical concentration is quite small but the temperature is quite high. However, in the laminar opposed flow diffusion flame configuration studied by Sarathy et al.[11], they did not find significant unimolecular decomposition in the fuel flow due to the low temperature inhibiting unimolecular decomposition. However, as the fuel approaches the flame zone, radicals are able to diffuse into the flow and thus cause abstraction to dominate.

For the case of premixed combustion, the presence of oxygen allows a radical pool to develop, which facilitates abstraction reactions. For this reason, in the JSR data of Dagaut et al.[49] and Sarathy et al.[11] there were very few products of unimolecular decomposition reactions.

However, at the high temperatures present in shock tube studies, Black et al.[54] and Moss et al.[9] found that unimolecular decomposition plays a role in fuel decomposition, despite having premixed fuel and oxidizer.

Several of these authors also produced kinetic schemes based on previous work. Sarathy et al.[11] proposed a mechanism based on the jet-stirred reactor and laminar flame speed data that is an extension of the work by Dagaut et al.[49,50]. Moss et al.[9] and Black et al.[54] created mechanisms based on their ignition delay measurements. Grana et al.[56] develop a mechanism based on a hierarchical approach and validate the mechanism with new non-premixed counterflow flame data, as well as flame and ignition data available in the literature. Finally, Harper et al.[58] included the data of Black et al.[54], Moss et al.[9], McEnally and Pfefferle[55] and Dagaut et al.[49], as well as an original pyrolysis experiment in their mechanism. The primary features of each mechanism related to ignition delay are discussed below.

Moss et al.[9] (referred to as the RPI mechanism) used the EXGAS[59] and THERGAS[60] software packages to create their mechanism and associated thermodynamics file. EXGAS is designed to automatically create a three-part reaction mechanism. The first part is a combination of the initial organic compounds and molecular oxygen. The second part involves all the reactions that consume the products of the first part but that are not included in the third part. Finally, the third part is a complete reaction base of $C_0$-$C_2$ reactions coupled with unsaturated $C_3$ and $C_4$ reactions. The mechanism agreed well with their data but was only validated with high-temperature and relatively low pressure studies.

The mechanism of Black et al.[54] (referred to as the NUI mechanism) also used EXGAS to create an *n*-butanol sub-mechanism that was linked to an updated $C_4$ mechanism from the same group. The sub-mechanism was then modified as needed. Black et al.[54] also performed *ab initio* calculations of the enthalpy of formation and bond dissociation energies for *n*-butanol. Their mechanism was also only validated against high-temperature shock-tube studies.

The mechanism of Grana et al.[56] (referred to as the Milano mechanism) was built hierarchically from low to high molecular weight compounds. The base of this mechanism is the $C_0$-$C_{16}$ semi-detailed mechanism created by the same group in several previous works. A new sub-mechanism for all four isomers of butanol is added to this set of base reactions. The mechanism is validated with species profiles in a non-premixed counterflow flame, as well as with species profiles available from other experiments in the literature. Ignition delay results are validated against the high-temperature shock-tube studies from Moss et al.[9] and Black et al.[54].

Finally, the mechanism of Harper et al.[58] (referred to as the MIT mechanism) was created by the software Reaction Mechanism Generator (RMG)[61,62] from MIT. This software creates a mechanism for given temperature, pressure and species conditions using a rate-based approach. To expand this initial mechanism, the software can be supplied with seed mechanisms that are included as the base for the full mechanism. The MIT mechanism includes GRI-Mech 3.0[63] and an ethanol mechanism by Marinov[64] as seed mechanisms. The first part of the full butanol mechanism was generated using the experimental conditions from the butanol-doped methane flames as reported by McEnally and Pfefferle[55]. This mechanism was then supplied as a seed mechanism for simulations of infinitely rich mixtures (i.e., pyrolysis) and very lean mixtures. This second iteration was used for simulations in CHEMKIN-PRO[65] duplicating experimental conditions used by Sarathy et al.[11], Moss et al.[9] and Black et al.[54]. Sensitivity analysis was performed on these simulations, and the resulting "important" reaction rates were improved using values from the literature or from quantum mechanical calculations. This mechanism has been validated across a wide temperature range, but the ignition-delay results have only been validated in the high-temperature regime.

Thus, a limited amount of work has been done on the fundamental kinetics of butanol combustion and, in particular, ignition at elevated pressures. A detailed chemical kinetic model should be able to reliably predict the emissions and combustion performance of butanol and to help determine its suitability as a replacement for gasoline. In formulating this model, it is important to include data from several different types of experiments over many temperature and pressure regimes. The goal of this study is to provide additional data relevant to engine conditions to be used in a reliable reaction mechanism for butanol, over regimes that have not been previously studied. Using a rapid compression machine (RCM), ignition delays for higher pressure conditions (7-30 bar) and low to intermediate temperature conditions (650-900 K) are obtained and modeled.

## 1.5     Organization of Thesis

The remainder of this thesis explains the experimental apparatus, modeling procedure and results of the investigation. Chapter 2 is devoted to the experimental apparatus; Chapter 3 describes the models used to simulate experimental results. Chapter 4 contains ignition-delay results, comparison to simulations and discussion. Finally, Chapter 5 summarizes the present findings as well as provides recommendations for future work.

## Chapter 2.  Experimental Methods

### 2.1  Experimental Apparatus

The rapid compression machine (RCM) used in the present work has been described elsewhere[66,67] and subsequently used to study the autoignition of *n*-decane, methylcyclohexane, syngas ($H_2/CO$), dimethyl ether, toluene and benzene[68-72]. The RCM consists of a reaction chamber, hydraulic cylinder, pneumatic cylinder and driving air tank (Figure 2.1). The reaction chamber is connected to the front of the hydraulic cylinder; the rear of the pneumatic cylinder attaches to the driving air tank. The reaction chamber has a bore of 2 inches, and the bore of the pneumatic cylinder is 5 inches, allowing a factor of 6.25 lower driving pressure than reaction chamber compressed pressure. The stroke of the RCM is variable from 7 to 10 inches by varying the number of spacers placed at the rear of the hydraulic cylinder. The reaction chamber can be fitted with split shims to allow variation of the piston clearance at top dead center (TDC).

Fuel/air pre-mixtures are prepared in a 17.5-L mixing tank. The mixing tank is connected to the reaction chamber by flexible stainless steel manifold tubing. The tank, reaction chamber and connecting manifold are wrapped in heating tape and insulation to control the initial temperature of the mixture. Temperature controllers from Omega Engineering use thermocouples placed on the lid of the mixing tank, approximately in the center of the mixing tank, embedded in the wall of the reaction chamber and near the inlet valve of the reaction chamber to control the preheat temperature of the mixture. A static pressure gauge (Omega Engineering, 0-50 psia) measures the pressure in the manifold and mixing tank.

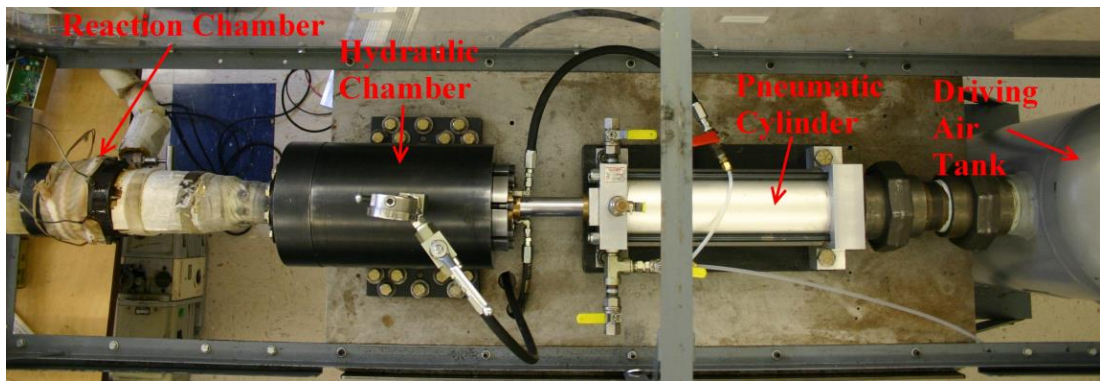

**Figure 2.1: The Rapid Compression Machine**

The piston in the reaction chamber is connected by a rod to pistons in the hydraulic cylinder and pneumatic cylinder. When retracted, the piston assembly is held in place by oil pressure on the front face of the piston in the hydraulic cylinder. The reaction chamber is loaded with a charge through a low-dead volume valve mounted on the side of the reaction chamber. The hydraulic pressure is released through a solenoid valve, and the piston is thrust forward by the pressure of the driving air on the rear face of the piston in the pneumatic cylinder. The pressure in the reaction chamber is measured by a dynamic pressure transducer (Kistler 6125B) with a charge amplifier (Kistler 5010B) and recorded by an NI-DAQ (PCI 6030E) at 25,000 samples per second. The reaction chamber can also be fitted with a rapid sampling apparatus, described below, or quartz windows to allow optical access.

One of the key features of the RCM is the ability to vary compressed temperature ($T_C$) while holding compressed pressure ($P_C$) constant. The initial pressure is varied to ensure that the compressed pressure remains constant as compressed temperature is varied. The compressed gas mixture temperature can be varied by changing the initial temperature of the fuel/air mixture, as well as by changing the compression ratio (i.e., by changing the stroke and clearance of the reactor piston). This allows a large range of conditions to be studied. For instance, the RCM in the present study has previously been used to reach compressed mixture pressures up to 45 bar and a compressed mixture temperature range from 600K to 1100K. In general, the compressed temperature and pressure are dependent on the compression ratio; this can be varied from about 5 to about 15.

## 2.2 Mixture Preparation

Fuel from Sigma-Aldrich (*n*-butanol, 99.9%) and gases from Airgas (O₂ 99.8%, Ar 99.998%, N₂ 99.998%) are used as the reactants. To determine the mixture composition, the mass of fuel, equivalence ratio and oxidizer ratio ($X_{O_2}:X_{inert}$, where $X$ indicates mole fraction) are specified.

To prepare a mixture, the tank is vacuumed to less than 5 torr and flushed with inert gas. The tank is vacuumed again to less than 3 torr, and the liquid fuel is injected by syringe through a septum. Since *n*-butanol is liquid at room temperature and has a relatively low vapor pressure (~6 torr at 25°C), it is measured gravimetrically in a syringe to within 0.01 g of the specified value. The mass of the syringe is measured again after injection to verify the mass of fuel in the mixing tank. Proportions of O₂, N₂ and Ar in the mixture are determined manometrically and added at room temperature, based on the pressures by the ideal gas law. The fuel vapor pressure is computed from correlations in the *Chemical Properties Handbook* by Yaws[73]. The preheat temperature is set above the saturation temperature of *n*-butanol for each mixture to ensure complete vaporization of the fuel. A magnetic stirrer mixes the gases in the tank. Finally, the heaters are turned on, and the temperature inside the mixing tank is allowed approximately 1.5 hours to reach steady state.

## 2.3 Check of Mixture Composition

Tests with GCMS are conducted to check that the expected mixture is present in the mixing tank for the entire duration of experiments. A mixture is prepared exactly as previously described, except a known concentration of *iso*-octane is added. This functions as an internal standard from which the concentration of butanol can be calculated. A 50cc sample bottle is heated to 150°C. The sample bottle has a two way valve and septum on one end to facilitate withdrawing samples from the bottle. The bottle is vacuumed before the mixture is admitted. The bottle is moved to the GCMS, where a sample is withdrawn with a Valco Gas-Tight syringe. This sample is injected into the split sampling valve on the GCMS. The sample is separated on a 100m long RTX-1 PONA column from Restek. The column temperature is held at 50°C for three minutes and then ramped at 4°C per minute to 100°C. The RTX-1 PONA column provides good separation of *n*-butanol and *iso*-octane, as shown in Figure (2.2a).

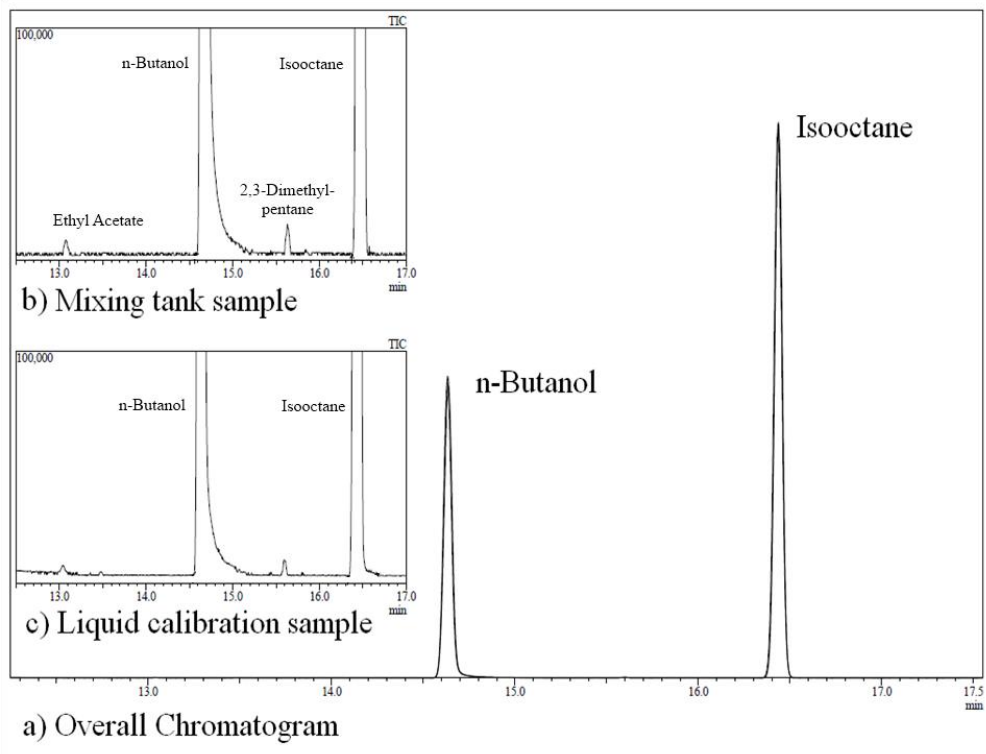

Figure 2.2: a) Overall chromatogram of *n*-butanol and *iso*-octane separation for $\phi = 0.5$ mixture, at mixing tank temperature of 87°C. b) Enlargement of a) from 12.5 to 17 minutes. c) Chromatogram of the liquid sample used for calibration. Comparison of b) and c) shows no decomposition of *n*-butanol in the mixing tank.

The first objective of the mixture composition check is to ensure there is no decomposition of *n*-butanol during preheating in the mixing tank. Figure (2.2c) shows the chromatogram of a liquid calibration sample consisting of *n*-butanol and *iso*-octane (Sigma-Aldrich, 99.9%) diluted in acetone (Sigma-Aldrich, 99.5%). This liquid sample was injected to the GCMS and the same temperature program described previously was applied. The two small peaks are impurities in the *iso*-octane and *n*-butanol; ethyl acetate is from *n*-butanol and 2,3-dimethylpentane is from *iso*-octane. Figure (2.2b) is an enlargement of Figure (2.2a) from 12.5 to 17 minutes. The mixing tank was heated to 87°C and contained a $\phi = 0.5$ mixture, with 1.1% by mole *iso*-octane replacing an equivalent amount of nitrogen. The *iso*-octane in the mixture functions as an internal standard. No additional peaks are present in Figure (2.2b) compared to Figure (2.2c), indicating there was no decomposition of *n*-butanol.

The second objective of the mixture composition check was to verify that the concentration of *n*-butanol matched the expected value. The expected value was calculated as the mole fraction of *n*-butanol in a $\phi = 0.5$ mixture in air. The response factor of *n*-butanol to *iso*-octane in the liquid sample was calculated by Equation (2.1):

$$\frac{\text{Peak Area of } n\text{-butanol}}{\text{Concentration of } n\text{-butanol}} = \text{Response Factor} * \frac{\text{Peak Area of } iso\text{-octane}}{\text{Concentration of } iso\text{-octane}} \qquad (2.1)$$

| | **Response Factor Determination** | | |
|---|---|---|---|
| | 1-butanol Peak Area | 10788294 | |
| | iso-octane Peak Area | 20061355 | |
| | ***Response Factor*** | **0.349076** | |
| | | | |
| | **Concentration of 1-Butanol** | | |
| | **Concentration of iso-octane in mixing tank** | | 0.0111 |
| *Sample #* | *1-butanol Peak Area* | *iso-octane Peak Area* | *Concentration of 1-butanol* |
| 1 | 11165856 | 22824035 | 0.015653 |
| 2 | 8920318 | 19145108 | 0.014908 |
| 3 | 18182941 | 35118170 | 0.016566 |
| 4 | 15224721 | 27259402 | 0.01787 |
| 5 | 19145886 | 34907200 | 0.017549 |
| | | **Average** | **0.0165** |
| | | **Expected** | **0.0172** |
| | | **Deviation (%)** | **4.05** |

**Table 2.1: Calculation of Response Factor and concentration of *n*-butanol for $\phi = 0.5$ in air mixture.**

Then, samples from the mixing tank were withdrawn using the procedure described above and analyzed using the GCMS. The peak area of each component was calculated, and using the known concentration of *iso*-octane and the response factor, the concentration of *n*-butanol was calculated. A total of five samples from one mixing tank were analyzed. The concentration of *n*-butanol was found to be within 4% of the expected value for this representative case. Table (2.1) shows the calculation of the response factor from the liquid sample in the top portion and the calculation of the concentration in the bottom portion.

| Mole Percentage (%) | | | | Equivalence Ratio | Pressure (bar) |
|---|---|---|---|---|---|
| n-Butanol | $O_2$ | $N_2$ | Ar | $\phi$ | $P_C$ |
| 2.96 | 20.39 | 76.65 | 0.00 | 0.87 | 30 |
| 2.96 | 20.39 | 0.00 | 76.65 | 0.87 | 7 |
| Mixture in air | | | | | |
| 3.38 | 20.30 | 76.32 | 0.00 | 1.0 | 15 |
| 3.38 | 20.30 | 76.32 | 0.00 | 1.0 | 30 |
| 1.72 | 20.65 | 77.63 | 0.00 | 0.5 | 15 |
| 6.54 | 19.63 | 73.83 | 0.00 | 2.0 | 15 |
| Vary $O_2$ Concentration | | | | | |
| 3.38 | 40.60 | 56.02 | 0.00 | 0.5 | 15 |
| 3.38 | 10.15 | 86.47 | 0.00 | 2.0 | 15 |
| Vary Fuel Concentration | | | | | |
| 1.69 | 20.30 | 78.01 | 0.00 | 0.5 | 15 |
| 6.76 | 20.30 | 72.94 | 0.00 | 2.0 | 15 |

**Table 2.2: Experimental conditions considered in this study.**

## 2.4 Experimental Conditions

Experiments are carried out under a wide variety of conditions. The pressure conditions chosen have not been covered by previous work; in addition, the pressure conditions are similar to those used in practical combustion devices. The fuel loading conditions were also chosen to cover ranges not studied in previous work. Most of the shock tube studies have used relatively dilute mixtures – only 1-3 percent by mole of fuel. This study includes conditions at twice the maximum fuel concentration of previous work. Furthermore, at equivalence ratios of 0.5 and 2.0, the concentrations of fuel and oxygen are separately adjusted to reveal the effect of each on the ignition delay. See Table (2.2) for the conditions studied in this work.

## 2.5 Experimental Reproducibility

Each compressed pressure and temperature condition is repeated at least 6 times to ensure reproducibility. The mean and standard deviation of the ignition delay for all runs at each condition are calculated; as an indication of reproducibility, the standard deviation is less than 10% of the mean in every case. Representative pressure traces for simulations and plotting are chosen as the closest to the mean. Figure (2.3) shows a set of runs from a representative condition. Furthermore, each new mixture preparation is checked against previously tested conditions to ensure consistency.

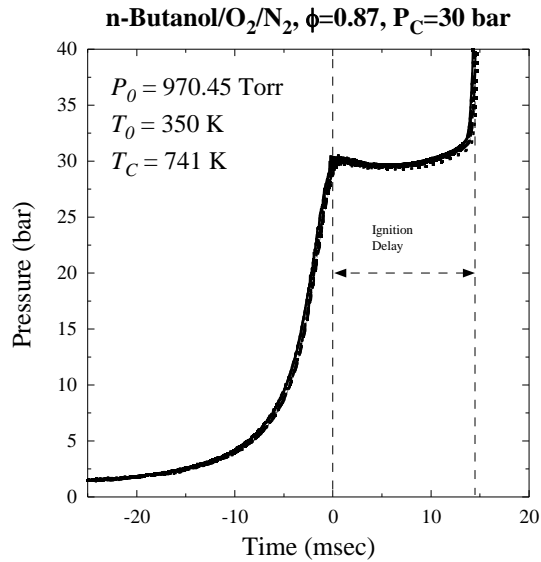

**Figure 2.3: Representative set of runs showing the reproducibility of experiments.**

## 2.6    Rapid Sampling Apparatus

The rapid sampling apparatus (RSA) is a device designed to rapidly extract the contents of the reaction chamber at a specified time and quench any reactions. The RSA is mounted on the end of the reaction chamber. It is equipped with an aluminum diaphragm that is punctured at the appointed time by an aluminum rod. This admits the gases in the reaction chamber to the RSA, where a volume expansion of approximately 50:1 quickly cools the gases. The rear of the sampling chamber is convex to reflect and weaken shock waves that occur when the diaphragm is punctured. The time of puncturing is controlled by a DG535 Digital Delay/Pulse generator.

The controlling circuit is an improvement of the design used in previous work[67]. The previous design used a spring to thrust the puncturing rod forward. Until the time of puncturing, the spring was compressed by an electromagnet. This design necessitated a complex circuit, and the electromagnet was prone to de-energizing before the appointed time. The new design attempts to correct this by using a device that does not have to be powered to hold the RSA in the non-punctured state. In this design, a push-type solenoid is attached to the puncturing rod. The solenoid is wired to a 12V power supply that is switched by a MOSFET. The MOSFET is controlled by the signal from the DG535, allowing current to the solenoid only when switched by the DG535. Thus, power to the entire circuit can be switched off until only a few moments before it is needed, whereas the previous design required voltage from the moment of installing the diaphragm until puncturing, which could be a long time.

Sampling results are very useful to provide speciation data. Speciation data can be used to verify the pathways in a reaction mechanism, which is an important part of creating a comprehensive reaction mechanism. Improving the puncture mechanism will make it much easier to quickly obtain speciation data and provide recommendations for modelers.

# Chapter 3. Simulations

## 3.1 Motivation for Simulations

In large part, the motivation for this study is to provide additional data for validation of chemical kinetic schemes. To that end, current reaction mechanisms are tested here against experimental data. Furthermore, limitations of the RCM require that the temperature at TDC be computed rather than measured. The procedure for this will be discussed in due course.

## 3.2 Types of Simulations

Two types of simulations, RCM and shock tube, are used in this study. Both types are modeled using CHEMKIN-PRO[65] and SENKIN[74] combined with CHEMKIN-III[75]. Although RCM simulations are preferred for comparison with the data in this study, shock tube simulations are also performed for comparison against previous work.

### 3.2.1 RCM Type

The first type of simulation is the RCM type. This model includes the compression stroke of the RCM by varying the volume of the simulated reaction chamber as a function of time. It also includes post-compression events such as heat loss of the reactants to the reaction chamber walls and ignition.

CHEMKIN-PRO and SENKIN include a method for varying the volume of the reaction chamber as a function of time. A custom subroutine for the RCM has been written with several variable parameters. First, the stroke and clearance of the piston must be specified. The primary goal of RCM type simulations is to match the simulated pressure trace to the experimental one during the compression stroke and the post-compression event prior to ignition. From experimental pressure traces, it is inferred that the piston initially travels with constant acceleration, followed by a period at constant velocity, before being decelerated to rest at a constant rate. Thus, the total compression time can be broken into three time periods, as shown in Eq. 3.1:

$$t_{comp} = t_{accel} + t_{cons} + t_{decel} \qquad (3.1)$$

where $t_{comp}$ is the total compression time, $t_{accel}$ is the acceleration time, $t_{cons}$ is the constant velocity time, and $t_{decel}$ is the deceleration time. In general, $t_{comp}$, $t_{accel}$, and $t_{decel}$ are chosen as parameters to model the compression stroke of the RCM, and Eq. 3.1 defines $t_{cons}$.

Heat loss from the reactants to the reaction chamber walls is handled separately during and after compression. During compression, heat loss is modeled by an additional volume that is added to the actual geometric volume of the reaction chamber. This is a user-set parameter called $V_{add}$.

Heat loss after compression is determined from a non-reactive experimental run. A non-reactive run is set up by replacing $O_2$ in the mixture with an equal amount of $N_2$ to maintain a similar specific heat ratio of the overall mixture. The heat loss after compression is modeled as an adiabatic volume expansion of the gases in the reaction chamber. This volume expansion is calculated from the experimental pressure trace, shown by Eq. 3.2:

$$PV^\gamma = const. \qquad (3.2)$$

where $P$ is the pressure, $V$ is the volume, and $\gamma$ is the temperature-dependent specific heat ratio. A polynomial fit to the calculated volume trace provides coefficients to model the heat loss after compression.

In summary, the model which provides the volume of the reaction chamber for RCM-type simulations including heat loss is given by Eq. 3.3. and 3.4:

$$\text{while } t \leq t_{comp}: V(t) = V_g(t) + V_{add} \tag{3.3}$$

$$\text{while } t > t_{comp}: V(t) = V(t_{comp}) * V_p(t) \tag{3.4}$$

where $V(t)$ is the volume of the reaction chamber, $V_g(t)$ is the geometric volume of the reaction chamber, $V(t_{comp})$ is the volume at $t_{comp}$ and $V_p(t)$ is the polynomial fit used to match the non-reactive, post-compression pressure trace.

The initial pressure for simulations is computed as a 5% pressure rise over the initial pressure in the experimental trace. This ensures that the pressure rise is due to compression and not electrical or mechanical noise. A pressure trace is computed and compared against the experimental pressure trace. The four parameters, $t_{comp}$, $t_{accel}$, $t_{decel}$, and $V_{add}$, are modified until a match is found.

### 3.2.2 Shock Tube Type

The second type of simulation is the shock tube type, which simulate conditions in the reaction chamber at the end of compression. This is accomplished by modeling the reaction chamber as a constant volume, adiabatic system. This method neglects heat loss of the reactants to the cold chamber walls and, as such, is useful for comparison with ignition delay results from shock tubes.

## 3.3 Determination of Temperature at TDC

Compression and ignition usually occur on a very short time scale compared to thermocouple response times. Thus, it is necessary to estimate the temperature of the compressed gases using numerical simulation. If the gas is compressed adiabatically, this can be done according to the relations:

$$\int_{T_0}^{T_{fc}} \frac{1}{\gamma-1} \frac{dT}{T} = \ln(CR) \tag{3.5}$$

$$\int_{T_0}^{T_{fc}} \frac{\gamma}{\gamma-1} \frac{dT}{T} = \ln\left(\frac{P_{fc}}{P_0}\right) \tag{3.6}$$

where $CR$ is the compression ratio, $T_{fc}$ and $P_{fc}$ are the fully adiabatic compressed temperature and pressure, respectively, $T_0$ and $P_0$ are the initial temperature and pressure, respectively, and $\gamma$ is the temperature dependent specific heat ratio. However, this does not account for the finite heat loss to the cold walls of the reaction chamber during compression. To permit calculation of the actual compressed temperature, it is assumed that heat loss only occurs in a thin boundary layer, and the majority of the gas is unaffected by heat loss. This is called the "adiabatic core" hypothesis and has been validated in Refs. [66, 68-72, 76]. As discussed previously, the reactor piston has been specifically designed to ensure an adiabatic core region in the reaction chamber.

Due to the heat loss, the realized compressed pressure ($P_C$) and temperature ($T_C$) are somewhat lower than the corresponding fully adiabatic case. However, adiabatic relations can still be used for the core region to determine the temperature there. Rather than using $T_{fc}$ and $P_{fc}$, the actual compressed pressure and temperature, $T_C$ and $P_C$, are used, as shown in Eq. 3.7:

$$\int_{T_0}^{T_C} \frac{\gamma}{\gamma-1} \frac{dT}{T} = \ln\left(\frac{P_C}{P_0}\right) \tag{3.7}$$

Direct integration of Eq. 3.7 is rather difficult as the specific heat ratio is a function of temperature. Thus, it is more practical to use SENKIN or CHEMKIN-PRO to perform RCM type simulations and determine $T_C$ using the measured pressure trace.

# Chapter 4. Results and Discussion

## 4.1 Data Analysis

Experimental data were analyzed in MATLAB to determine the ignition delay. Figure (4.1a) shows a typical pressure trace in the neighborhood of TDC, before a moving averaging scheme was applied to smooth high-frequency noise. Figure (4.1b) shows the smoothed pressure trace. Following the smoothing process, the derivative of the pressure was calculated by the second forward difference approximation. The end of compression, when the piston reached TDC, was found by the maximum of the pressure trace up to the ignition point. The maximum of the derivative after TDC was found, to determine the inflection point in the pressure trace, and this defined the point of ignition. The ignition delay was the time difference between the point of ignition and the end of compression. Figure (4.2) illustrates the definition of ignition delay ($\tau$) used in this study.

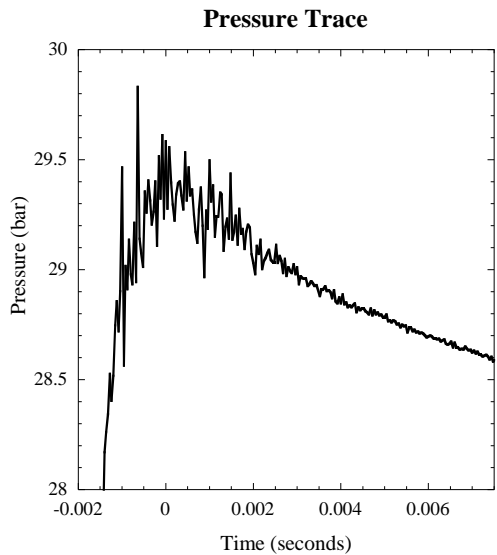

**Figure 4.1a: Raw Pressure Trace**

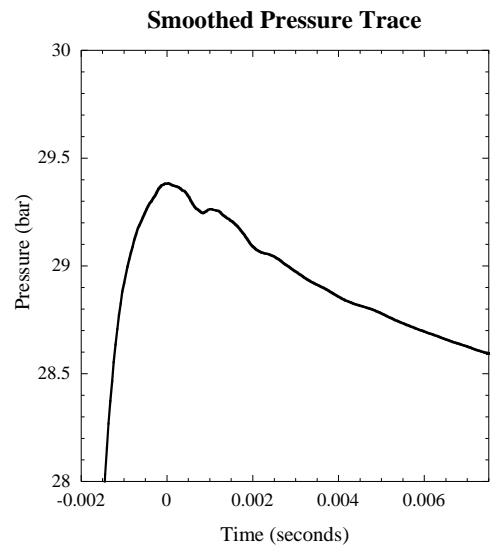

**Figure 4.1b: Smoothed Pressure Trace**

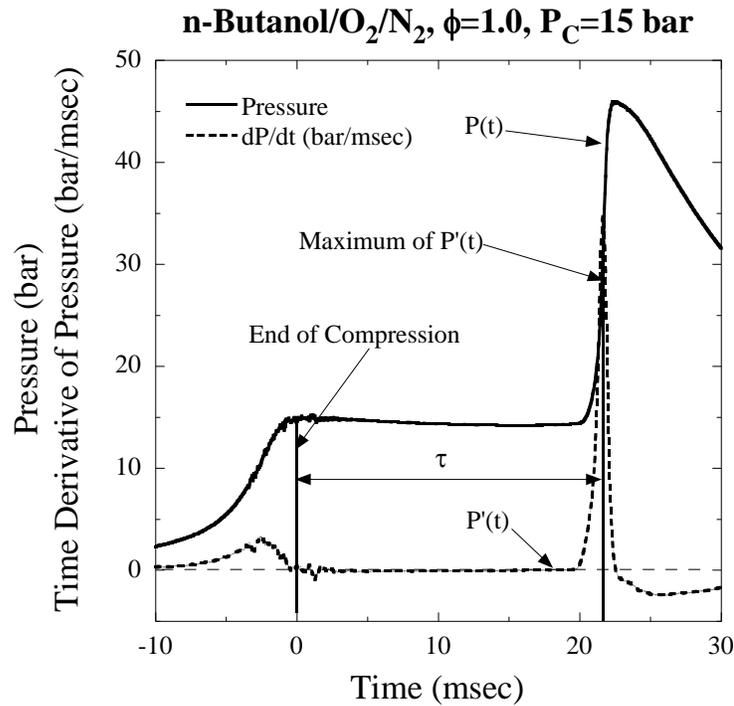

**Figure 4.2: Definition of ignition delay used in this study**

## 4.2 Experimental Results

Figure (4.3) shows one of the key features of the RCM, namely, the ability to vary compressed temperature at constant compressed pressure, as discussed in Chapter 2. The conditions in this figure are representative of conditions in all the experiments. As seen in Figure (4.3), ignition delay decreased monotonically as temperature increased, indicating that these experiments were not in the negative temperature coefficient (NTC) region. Indeed, no NTC region was observed for any conditions in this study. This was somewhat surprising, as *n*-butane exhibits a clear negative temperature dependence in a similar temperature and pressure range[77]. Propane also exhibits a negative temperature dependence in a similar temperature and pressure range[78], but the autoignition of *n*-propanol has not been studied in the present temperature and pressure range to determine if a comparison can be drawn[79,80].

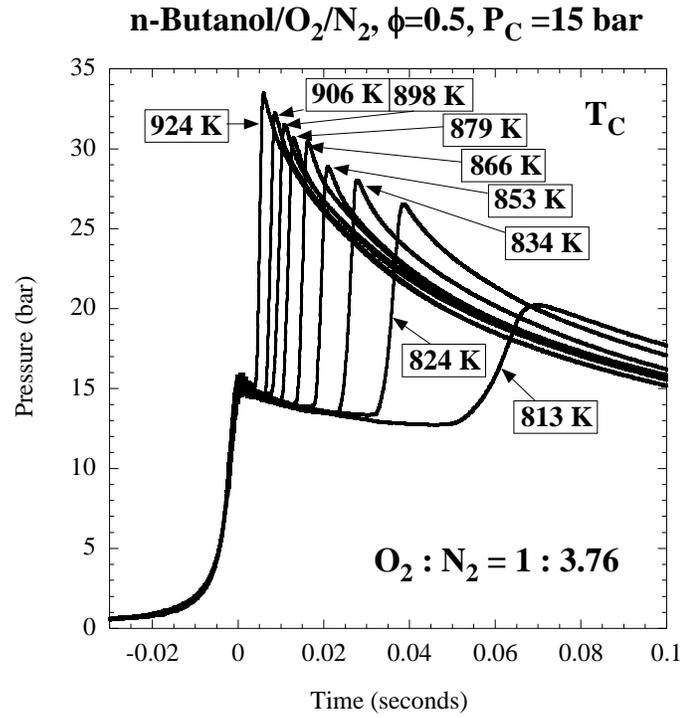

**Figure 4.3:** Pressure traces for the $\phi = 0.5$ in air experiments.

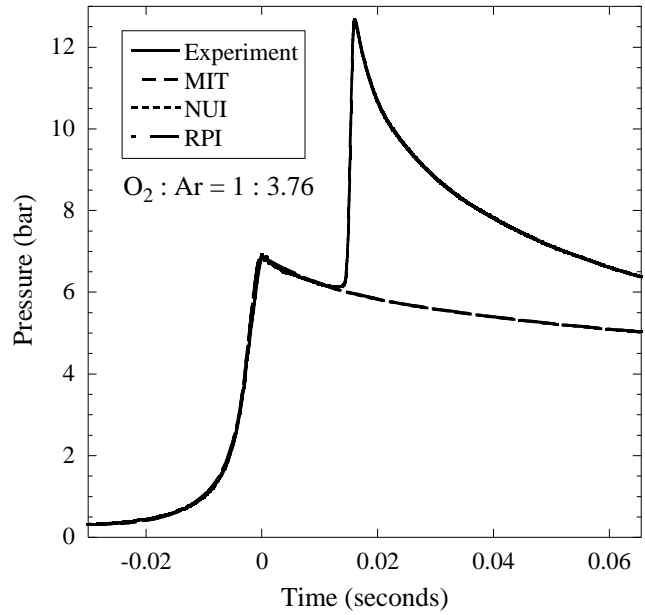

**Figure 4.4:** Comparison of RCM simulations with experimental data

It is also clear from Figure (4.3) that two-stage ignition did not occur. This was the case for all the conditions studied in this work and agrees with the work of Zhang et al.[43]. However, $C_3$ and greater alkanes exhibit two-stage ignition in the temperature and pressure range studied[77,81]. Again, *n*-propanol has not been studied in this temperature and pressure range to determine if a comparison is suitable. Finally, although Cullis and Warwicker[53] reported cool flames of *n*-butanol at low temperatures (575K-625K), their work was also at very low pressure (100-240 torr) and, therefore, is not directly comparable to the current, high-pressure, results.

Figure (4.4) shows a comparison of RCM-type simulations with experimental data at a representative condition. For all conditions in this study, none of the mechanisms were able to reproduce the ignition delay within the experimental duration, despite matching the pressure trace during the compression stroke and post-compression, pre-ignition event very well. Therefore,

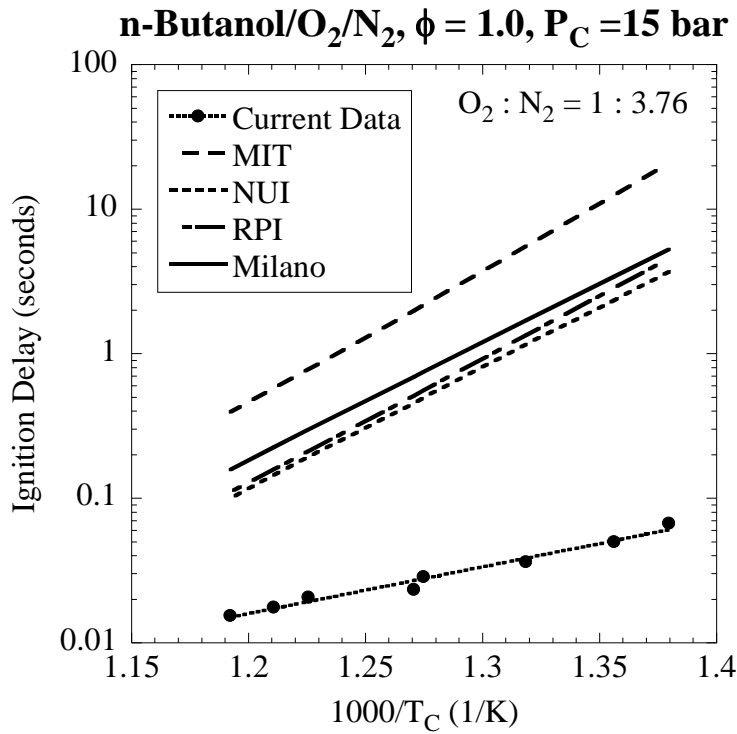

**Figure 4.5: Comparison of computed ignition delays using shock tube simulations and experimental results. The line through the experiments is a least squares fit to the data.**

shock tube-type simulations are shown in the other figures in this work.

As shown by Figure (4.5), shock tube simulations of ignition delay were much longer than the experimentally measured ignition delays. It is noted that none of the mechanisms have been validated for ignition delay in this temperature regime, and the MIT mechanism over-predicts the ignition delay by more than the other mechanisms. However, due in part to the work performed here, the MIT group has realized an error in their reaction mechanism. They will publish their updated mechanism in an upcoming article[82]. The nature of the error will be further discussed in Section 4.3. It appears from Figure (4.5) that the simulations and experiments converge as the temperature increases. To improve clarity of the following graphs, simulation results have been omitted except where noted.

Figure (4.6) shows a comparison of three equivalence ratios made in air. Error bars are

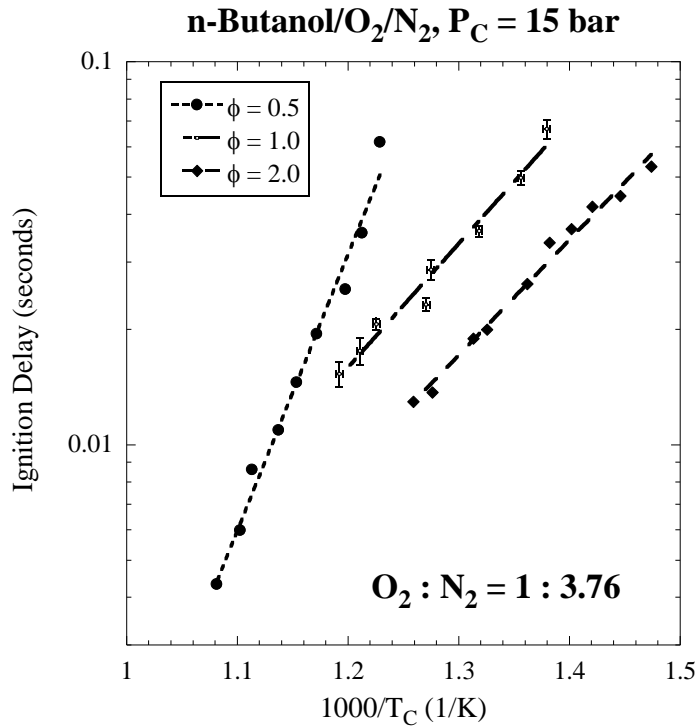

**Figure 4.6: Comparison of 3 equivalence ratios in air. Lines are least squares fits to the data.**

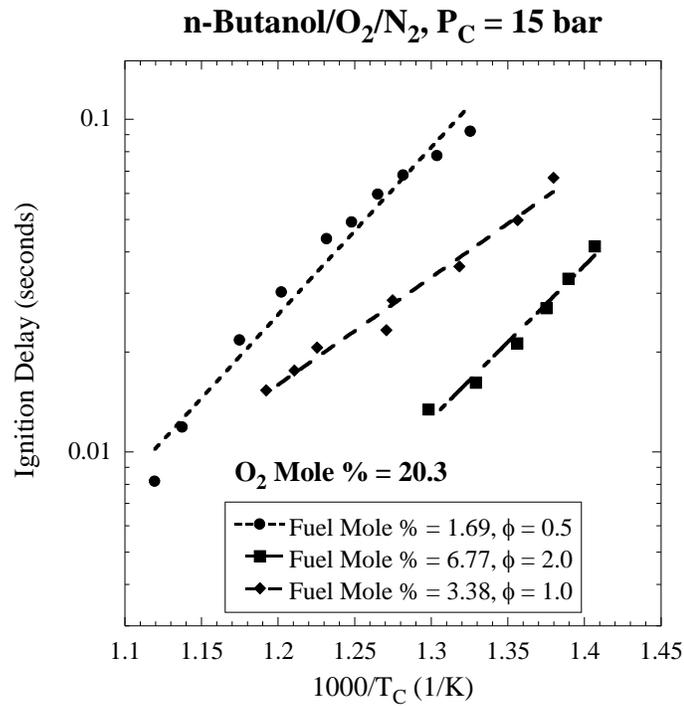

**Figure 4.7:** Comparison of three fuel concentrations at three equivalence ratios. In these, the oxygen mole fraction is held constant. Lines are linear least squares fits to the data.

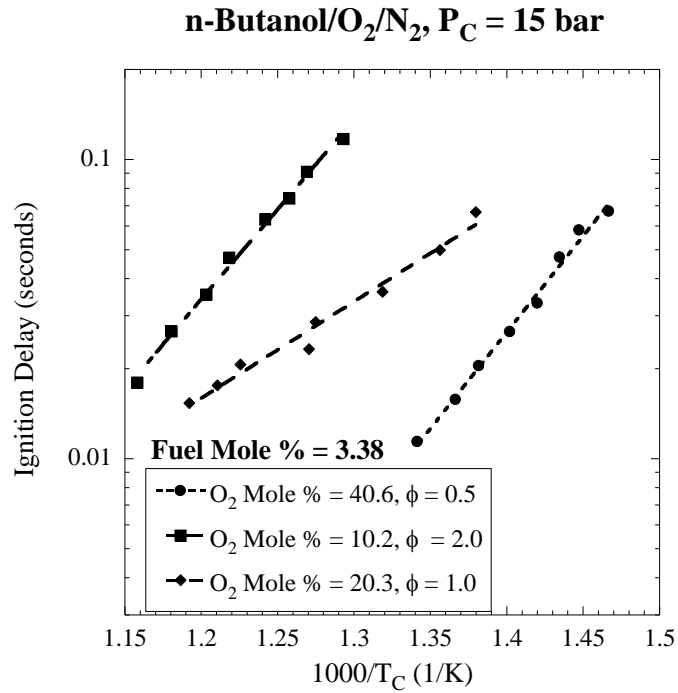

**Figure 4.8:** Comparison of three oxygen concentrations at three equivalence ratios. In these, the fuel mole fraction is held constant. Lines are linear least squares fits to the data.

shown for the $\phi = 1.0$ case. The vertical error bars are two standard deviations of the ignition delay. The standard deviation is computed based on all the runs at a particular compressed temperature and pressure condition. The horizontal error bars are the uncertainty in the measurement of the initial temperature, according to the manufacturer's specification. According to Ref. [69], this corresponds closely to the uncertainty in the calculation of the compressed temperature.

The effect of equivalence ratio on the ignition delay is quite clear – increasing the fuel concentration increased the reactivity in the temperature range investigated. That is, similar ignition delays were found at lower temperatures as the fuel concentration increased. This observation is corroborated by the results shown in Figure (4.7), which shows ignition delay as the fuel mole fraction varied with constant oxygen mole fraction. Again, the lowest fuel mole fraction was the least reactive, and increasing the fuel mole fraction increased reactivity. A similar effect of oxygen concentration is shown in Figure (4.8). Increasing the oxygen concentration increased reactivity at constant fuel mole fraction. It is further noted from Figure (4.6) that for butanol/air mixtures the reactivity dependence on the equivalence ratio can differ in different temperature ranges, as the lean mixture can become more reactive at high temperatures.

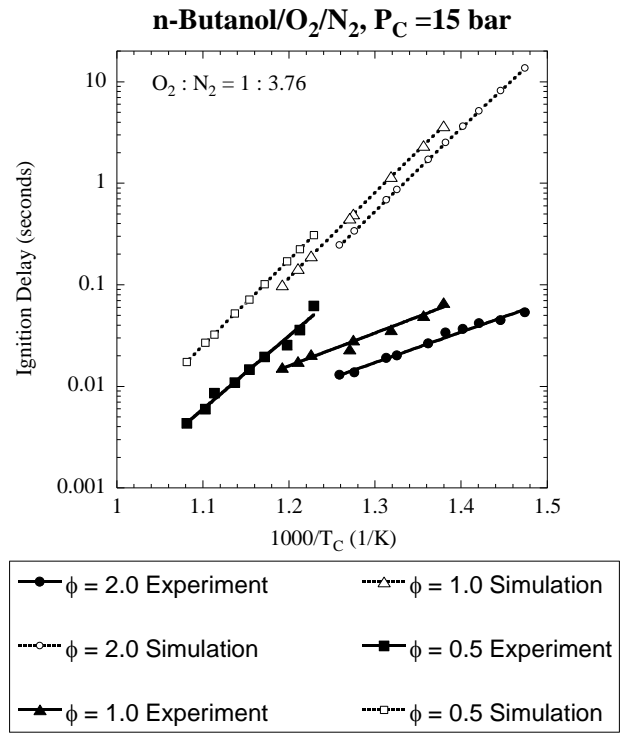

**Figure 4.9: Comparison of 3 equivalence ratios with simulations using the NUI mechanism. Simulations are dotted lines and open symbols. Filled symbols and solid lines are experimental data.**

Figure (4.9) shows a comparison of three equivalence ratios in air with the simulations using the NUI mechanism. The simulations are able to reproduce the difference between the equivalence ratios, in that they become more reactive as the equivalence ratio increases. However, the experiments at the $\phi = 1.0$ and 2.0 conditions have a much shallower slope than the simulations and the simulations for these two cases are two orders of magnitude longer than the experiments in the lowest temperature case. In terms of slope, the $\phi = 0.5$ case is relatively well reproduced. However, the simulations still over-predict the ignition delay by approximately

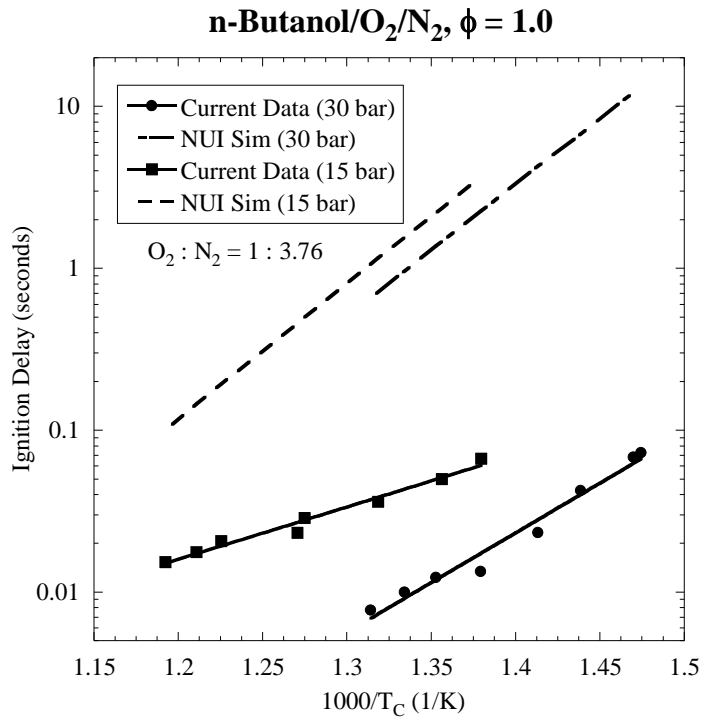

**Figure 4.10: Comparison of experiments at 15 and 30 bar for $\phi = 1.0$ in air.**

an order of magnitude.

Figure (4.10) shows comparison of data at 15 and 30 bar and $\phi = 1.0$ in air. The dashed lines are simulations using the NUI mechanism, while the solid lines through the experiments are least squares fits to the data. The 30 bar cases are more reactive than the 15 bar cases, as is expected. However, it is clear from the figure that the simulations are unable to reproduce the difference between the 15 and 30 bar cases. This suggests that the mechanism is not properly capturing the pressure dependence in this temperature range.

A non-linear regression was performed in MATLAB to determine a correlation of ignition delay with reactant mole fraction and compressed pressure. This correlation is given in Eq. 4.1:

$$\tau = 10^{-(8.5\pm.08)} * X_{O_2}^{-(1.7\pm0.2)} * X_{n-Butanol}^{-(1.4\pm0.2)} *$$

$$P_C^{-(1.5\pm0.3)} * \exp[(9730.3 \pm 1035.5)/T_C] \text{ (seconds)} \qquad (4.1)$$

where $X_n$ is the mole fraction of species $n$, and the uncertainties in the correlation parameters are the uncertainties in the regression resulting from scatter in the data. The result of this regression analysis is shown in Figure (4.11). This correlation is calculated over a relatively small range of temperature and does not include data from the literature at higher temperature and lower pressure. Thus, extending the correlation to temperature or pressure ranges far outside the current work is unlikely to yield acceptable results.

## 4.3    NUI Mechanism Analysis

As shown in Figures (4.5, 4.9 and 4.10), simulations for every experimental condition over-predict the ignition delay quite substantially. This is due to the fact that none of the mechanisms have been validated with low temperature/high pressure data. However, the degree of difference between simulations and experiments is somewhat surprising. A more detailed analysis of the NUI mechanism was carried out to determine the reasons for the discrepancy between simulations and experiments.

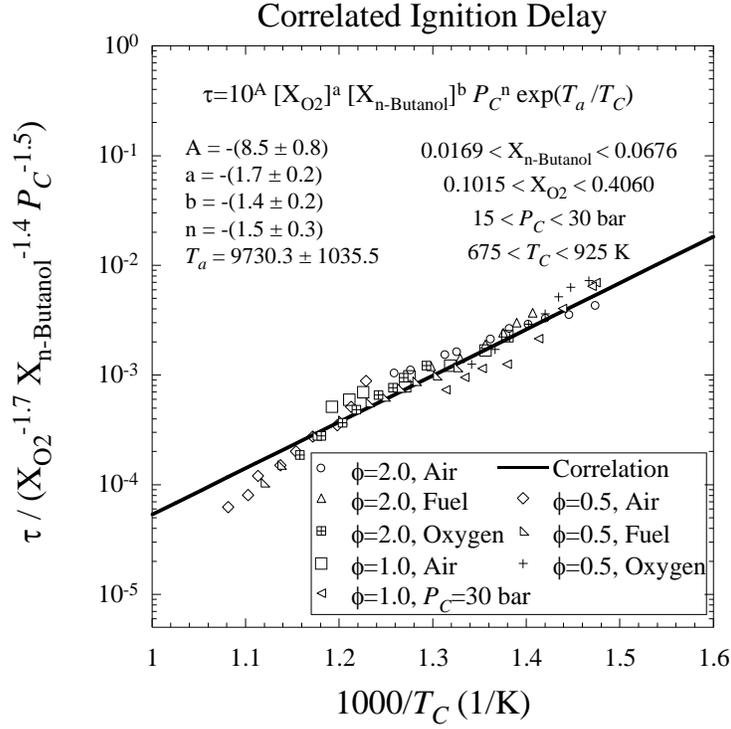

**Figure 4.11: Result of correlation given in Equation (4.1)**

### 4.3.1  Brute Force Sensitivity Analysis

First, brute force sensitivity analysis was conducted. Percent sensitivity is defined by Eq. (4.2):

$$\% \text{ sensitivity} = \frac{\tau(2k_i)-\tau(k_i)}{\tau(k_i)} \times 100\% \qquad (4.2)$$

where $k_i$ is the rate of reaction $i$, $\tau(2k_i)$ is the ignition delay when the rate of reaction $i$ has been doubled, and $\tau(k_i)$ is the nominal value. Thus, a positive value for sensitivity means that the ignition delay became longer when the rate of reaction $i$ was doubled. Figure (4.12) shows the results of this analysis, at 15 atm and $\phi = 1.0$ in air, over a range of compressed temperatures. Three initial temperatures were chosen: 1100K and 1800K correspond to the upper and lower temperatures for which this mechanism has been validated, while 700K is representative of the conditions in this study.

Figure (4.12) shows the ten most sensitive reactions from the 1100K case, along with the sensitivity of the same reactions at 700K and 1800K. Key radicals such as O, H, OH, $HO_2$ and $H_2O_2$ are important in ignition processes of hydrocarbon fuels[83]. Therefore, it is not surprising to see many reactions among those radicals in the results. The other reactions, however, involve the initial decomposition of the fuel. At low temperature, the system is clearly most sensitive to reaction 1353:

$$NC4H9OH + HO2 \Leftrightarrow C4H8OH\text{-}1 + H2O2 \qquad (R1353)$$

which is H-abstraction from the $\alpha$-carbon of 1-butanol. Figure (1.1) is reproduced here for reference to the structure of *n*-butanol. It is in this type of reaction (i.e. H-abstraction from the fuel by $HO_2$) where the error in the MIT mechanism lies. The rate-estimation database in their mechanism generation software, RMG, was providing rates which were much too slow (roughly 6 orders of magnitude in the temperature range of interest), thus causing the large discrepancy between simulated results and experimental results. It should be noted that the correction applied by the MIT group affects all reaction mechanisms for other fuels generated by their RMG software, not just this *n*-butanol one.

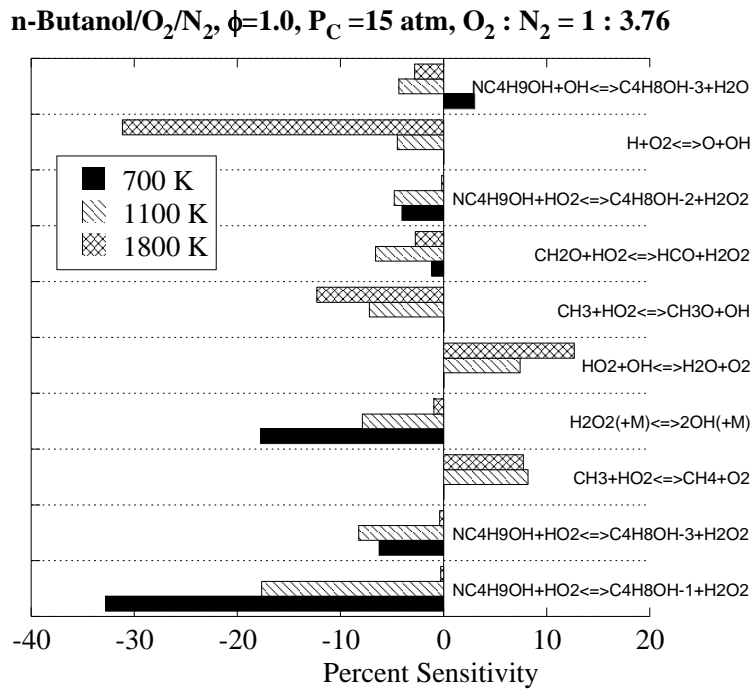

**Figure 4.12: Results of brute force sensitivity analysis.**

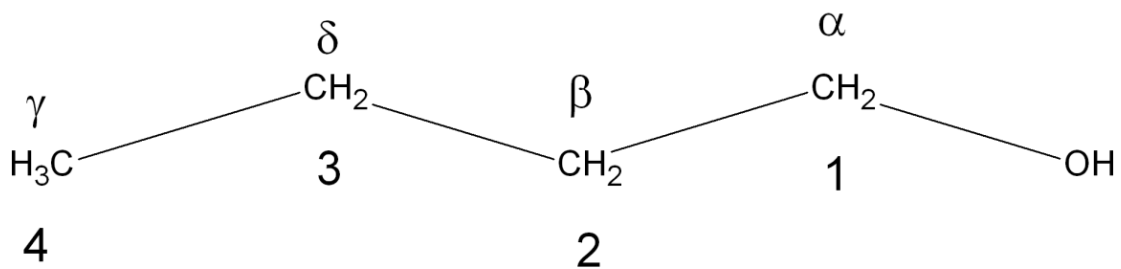

**Figure 1.1: Structure of *n*-Butanol**

At low temperatures, H-abstraction from the fuel plays a major role in the combustion process[81,83]. Zhang et al.[43] also showed that H-abstraction is important in low temperature 1-butanol ignition. Due to the effect of the alcohol group, the $\alpha$-carbon has the most weakly bonded hydrogens in 1-butanol[54]. Therefore, it is reasonable to expect the system to be very sensitive to R1353 at low temperature. By increasing the temperature, the sensitivity of the system to this reaction is reduced, such that, at 1800K, the sensitivity is less than 1%. This indicates that increasing the rate of this reaction may improve simulation results at low temperatures but will not strongly affect correlation with existing high temperature data for ignition delay.

A similar analysis was conducted at 800K while varying the pressure from 1 to 30 atm. The results of this analysis showed that the sensitivity of most reactions was similar over the entire range, suggesting that changing the rate of R1353 will not negatively impact correlation with previous results.

To determine the effect of increasing the rate of R1353, the ignition delay using the shock tube procedure was computed at several multiples of the reaction rate and two temperatures. The ignition delays from this procedure are shown in Figure (4.13). The temperature of 758K represents an experimental condition, while the 700K is the temperature of the previous sensitivity analysis. When the rate is increased more than approximately 500 times of the original value, the ignition delay is no longer reduced. For comparison with experimental results, R1353 was multiplied 1000 times its initial value. Shock tube simulations were calculated at the conditions of the 15 bar, $\phi = 1$ experiments. The results are shown in Figure (4.14). As expected, increasing the rate of R1353 brought the simulations much closer to the experimental results.

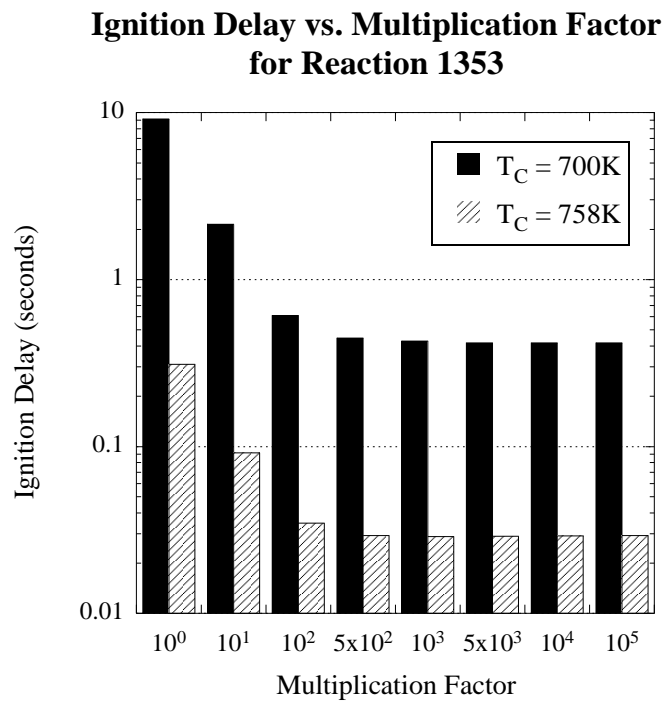

**Figure 4.13: Ignition delay as a function of multiplication factor for reaction 1353. Initial conditions are indicated in the legend. Both are $\phi = 1.0$.**

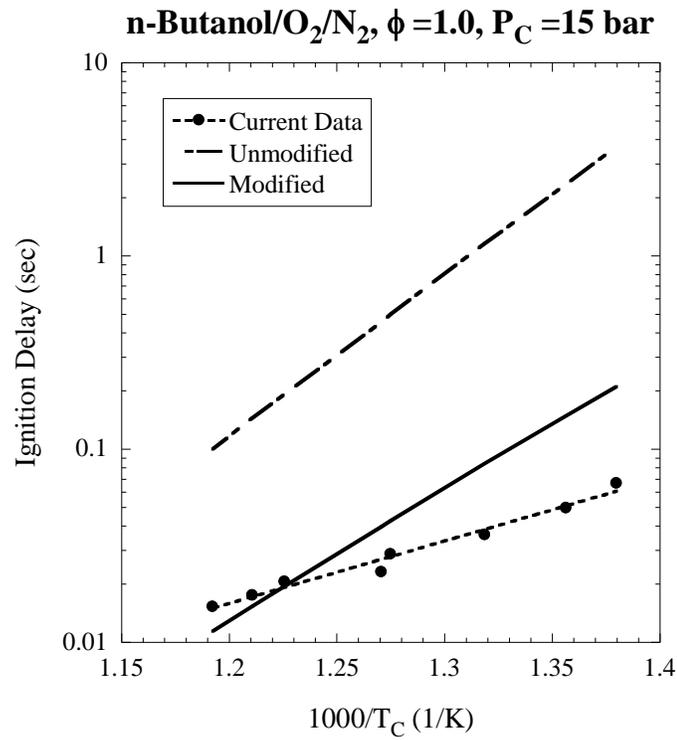

**Figure 4.14: Comparison of experimental ignition delays with the modified and unmodified NUI mechanism**

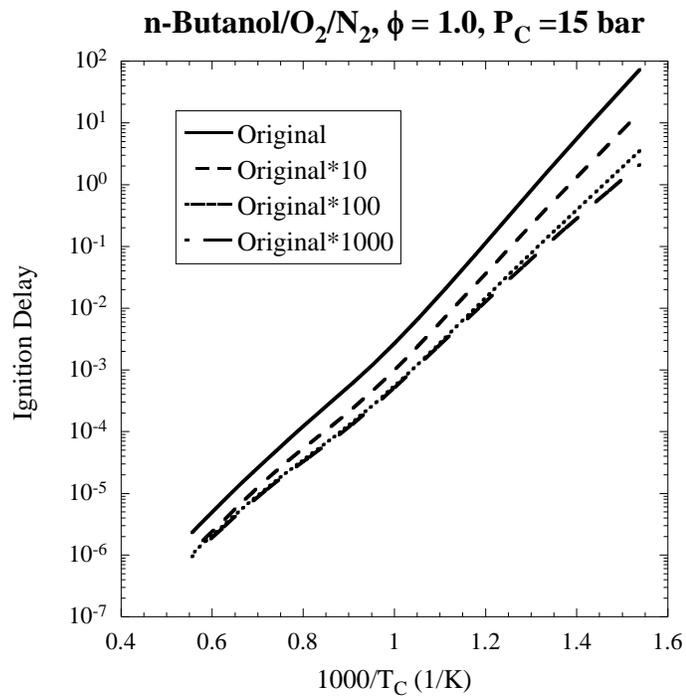

**Figure 4.15: Comparison of computed ignition delays over a large temperature range, for various multiplication factors of reaction 1353**

Nevertheless, at the lowest temperature, the difference between the results of simulations and experiments was still more than 100 ms.

Finally, results over a large temperature range were computed for the original mechanism and the modified mechanism. It can be seen in Figure (4.15) that even for low multiplication factors, the ignition delay becomes shorter over the entire temperature range. However, agreement with low temperature experiments is improved significantly as the reaction rate is increased, while agreement at high temperature is not significantly decreased.

### 4.3.2 Reaction Path Analysis

The second method of mechanism analysis is reaction path analysis. A reaction path diagram is one way of visualizing this analysis. The reaction path diagram shows the percent of each reactant destroyed to form the product indicated by the arrow. The percent destruction represents the cumulative destruction of each reactant up to the point in time where the mole fraction of the fuel is reduced 20% compared to the initial mole fraction. In this work, the reaction path diagrams begin with the fuel, *n*-butanol, located in the center of each diagram.

The time of 20% fuel consumption was chosen for two reasons: first, it has been used previously in the literature, and second, it is before the point of thermal runaway when the chemistry changes regimes. The first few levels of fuel decomposition are shown in Figure (4.16) for $\phi = 1.0$ in air, 15 atm and 800K. For comparison, Figure (4.17) shows the same species at $\phi = 1.0$ in air, 15 atm and 1600K. Note that the following analysis applies to the unmodified mechanism.

One of the most noticeable differences is the complete lack of unimolecular decomposition reactions, such as C-C bond fission of the $\alpha$-bond, in the low temperature case. These reactions are characterized by rather high activation energy, so it is not surprising that they are not a factor at low temperature. Also not surprising is the fact that H-abstraction from the $\alpha$- carbon consumes the largest percentage of fuel, as discussed previously. In fact, results do not become surprising until the second level of reactions – that is, reactions of the initially formed radicals. At this level, the primary species formed are unsaturated alcohols (also called enols because of the double bond in the carbon skeleton), such as but-1-en-1-ol, but-2-en-1-ol and but-3-en-1-ol. But-1-en-1-ol, for instance, is formed from the $\alpha$-hydroxyalkyl and $\beta$-hydroxyalkyl radicals by reaction with molecular oxygen. The $\beta$-hydroxyalkyl and $\gamma$-hydroxyalkyl radicals are each capable of forming the other two enols, depending on where the second hydrogen is removed.

In the 800K path diagram (Figure (4.16) and expanded in Figure (4.18)), it can be seen that nearly 30% of 1-butanol enters the but-1-en-1-ol pathway. The next step of this pathway is the removal of an additional H-atom to form $C_4H_6OH$, which can form a resonantly-stabilized structure. This is primarily destroyed back into the enol by $HO_2$. This step seems unlikely to be prominent, as it involves the release of an oxygen molecule. Indeed, at high temperature, this reaction is only a minor contributor, and the direction has been reversed; that is, the reaction is producing the resonant structure from the enol instead of the other way around. Part of this may have to do with the fact that $HO_2$ reactions are much less important at high temperature; however, it is useful to emphasize that the reaction direction has been reversed.

In the low temperature case, the final step of this pathway produces a diene, $C_4H_5OH$ (but-1,3-en-1-ol). This species accumulates in the system up to the 20% fuel consumption point. That is, the rate of production from each reaction is greater than the rate of destruction by each reaction for all the reactions involving this species. In addition, in this mechanism, this species can only be formed from $C_4H_6OH$ by spontaneous emission of a hydrogen atom. However, according to Walker and Morley[84], the primary method of formation of dienes from 1-butene and 2-butene is by reaction with molecular oxygen, to form the diene and $HO_2$. This pathway is supported by Vanhove et al.[85] who discussed general allyl decomposition pathways. They also noted the additional pathway of molecular oxygen addition followed by several isomerization steps, leading to the diene and $HO_2$.

Figure 4.16: Initial fuel decomposition in the NUI mechanism at initial conditions of 800K, 15 atm, $\phi = 1.0$

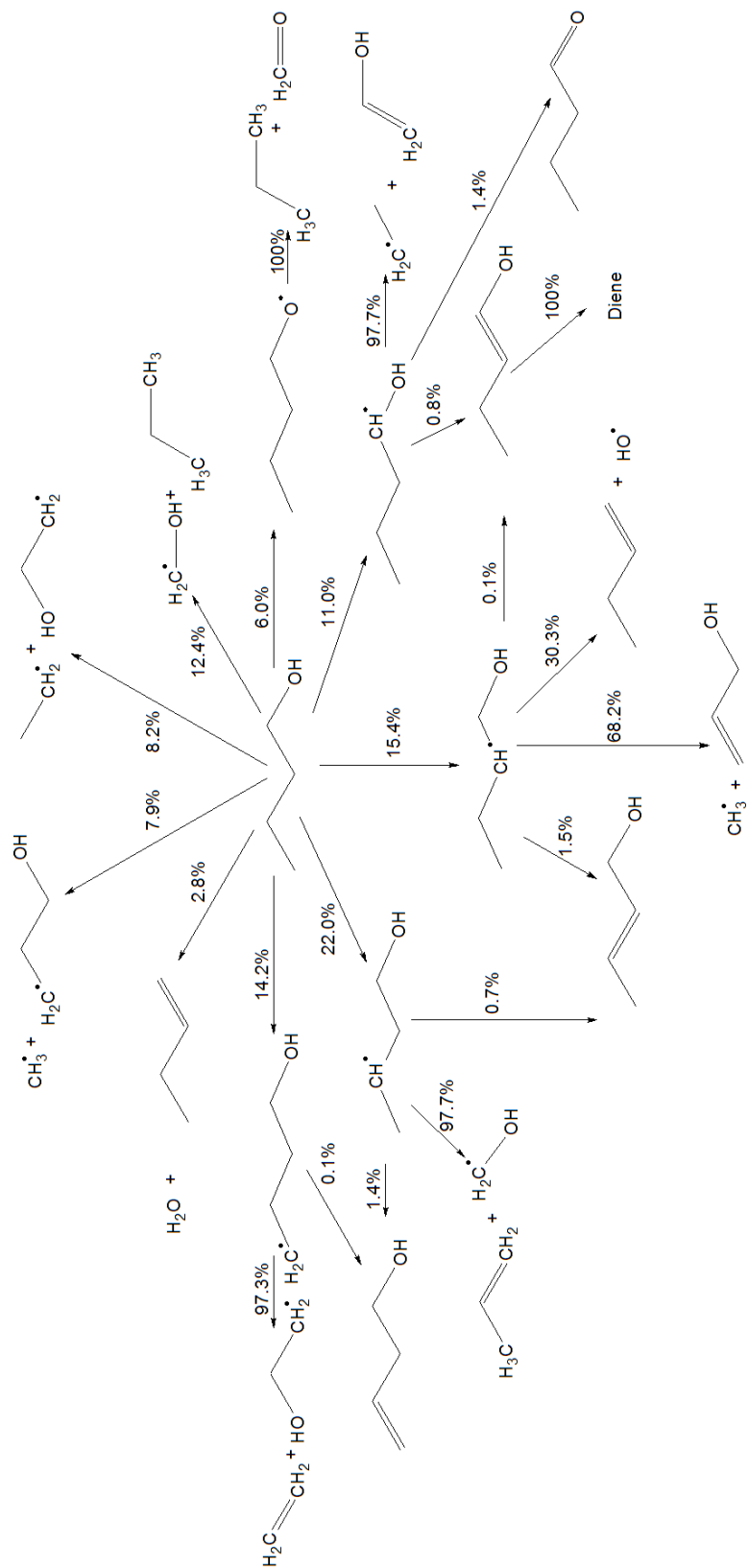

**Figure 4.17:** Initial fuel decomposition in the NUI mechanism at initial conditions of 1600K, 15 atm, $\phi = 1.0$

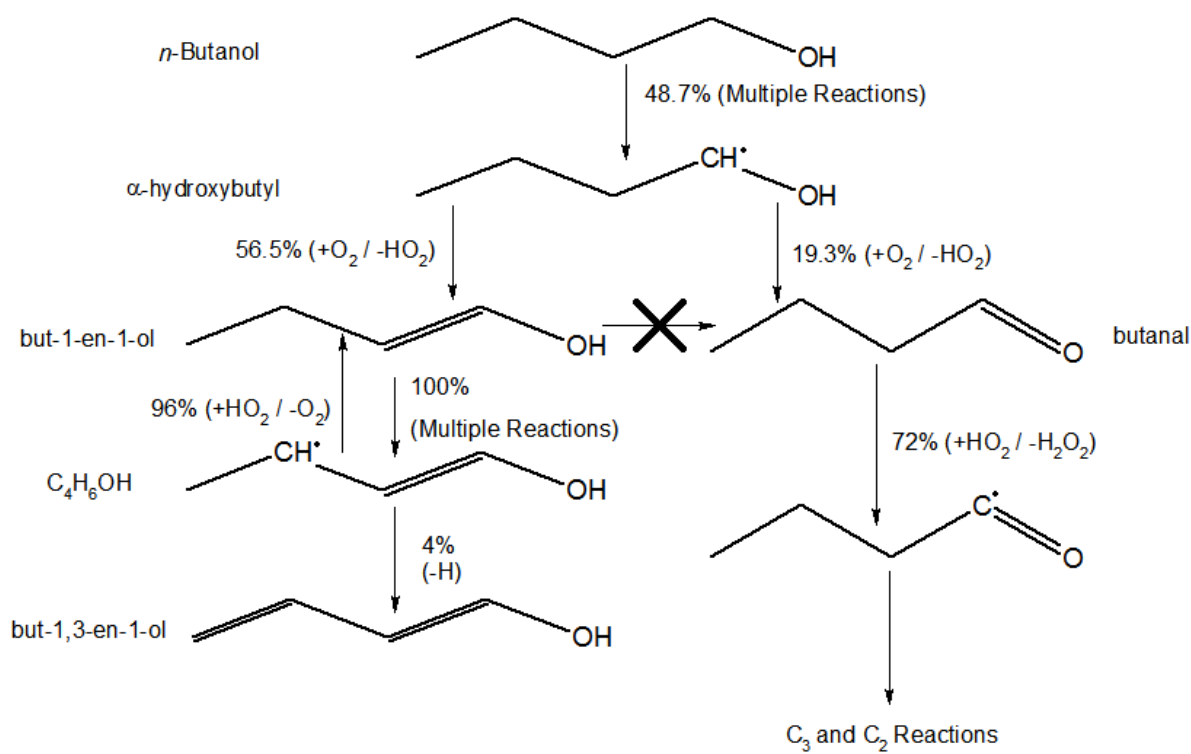

**Figure 4.18: Pathway of but-1,3-en-1-ol formation, through but-1-en-1-ol. Also, the pathway of butanal formation. The X through the arrow from but-1-en-1-ol indicates that pathway is not present in the mechanism.**

Another intermediate that seems to be important is butanal. Butanal has been detected by several studies, including flames[52], jet-stirred reactors[11] and engine studies[43]. It can be formed by several pathways, detailed in the work of Zhang et al.[43] and Grana et al.[56], and reproduced here for reference as shown in Figure (4.19). It can be seen from Figure (4.18) that, in the NUI mechanism, the formation of butanal occurs primarily by attack of molecular oxygen on the alcohol group of $\alpha$-hydroxybutyl radicals, producing butanal and $HO_2$. The pathway of tautomerization of but-1-en-1-ol to butanal is not present in this mechanism.

Black et al.[54] reported that the predicted concentrations of but-1-en-1-ol were much higher than butanal. However, experimental results from the jet-stirred reactor[49] showed that the concentration of butanal was higher than the simulated profiles of both but-1-en-1-ol and butanal, and was in fact approximately equal to the sum of the simulated butanal and but-1-en-1-ol profiles. Black et al.[54] postulated that rapid tautomerization of but-1-en-1-ol to butanal occurred between the reaction chamber and the analysis device, thus leading to the anomalous readings.

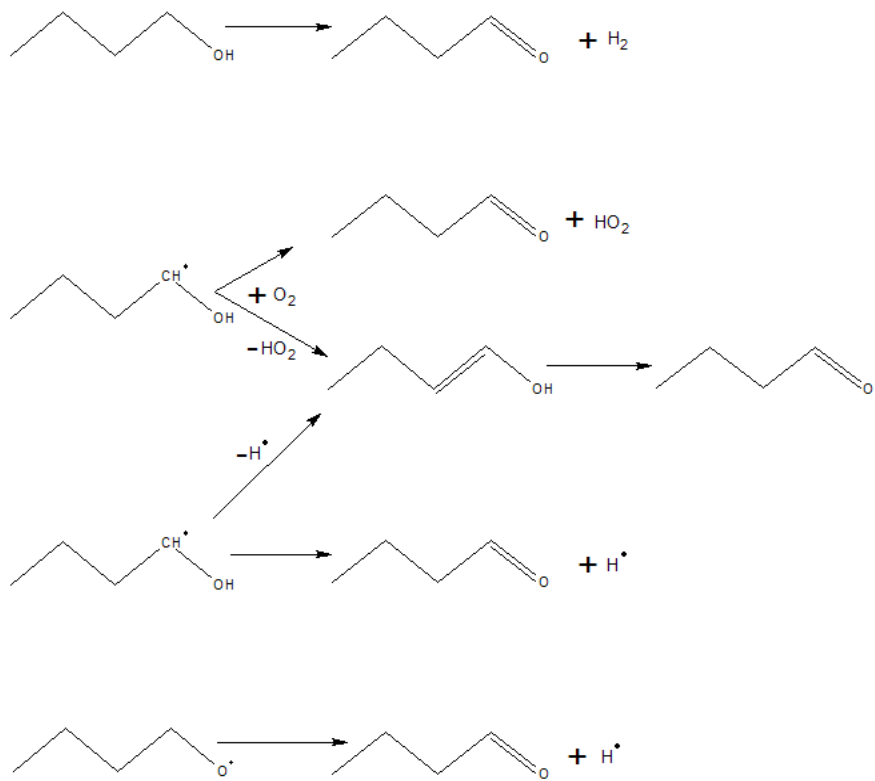

**Figure 4.19: Pathways of butanal formation. From Ref. [43]**

In contrast, Grana et al.[56] proposed that, although enols will be formed at high temperatures, tautomerization is so rapid that only the corresponding ketones will be present. This approach does not seem to agree with the results of the study by Yang et al.[52] who reported ionization signals at the ionization threshold of all of the butenols, propenols and ethenol. Other works in the literature show that, although enols are a minor species in most flames, they do exist and need to be accounted for[86].

Finally, Zhang et al.[43] reported small concentrations of but-2-en-1-ol and but-3-en-1-ol and 300 times higher concentrations of butanal in their engine ignition study. They thus infer that butanal was primarily formed directly from $\alpha$-hydroxybutyl, rather than primarily by tautomerization from but-1-en-1-ol. In any case, the reactions of the $\alpha$-hydroxybutyl radical, both formation and destruction, are apparently among the most important in the system, and a thorough investigation of these pathways is warranted.

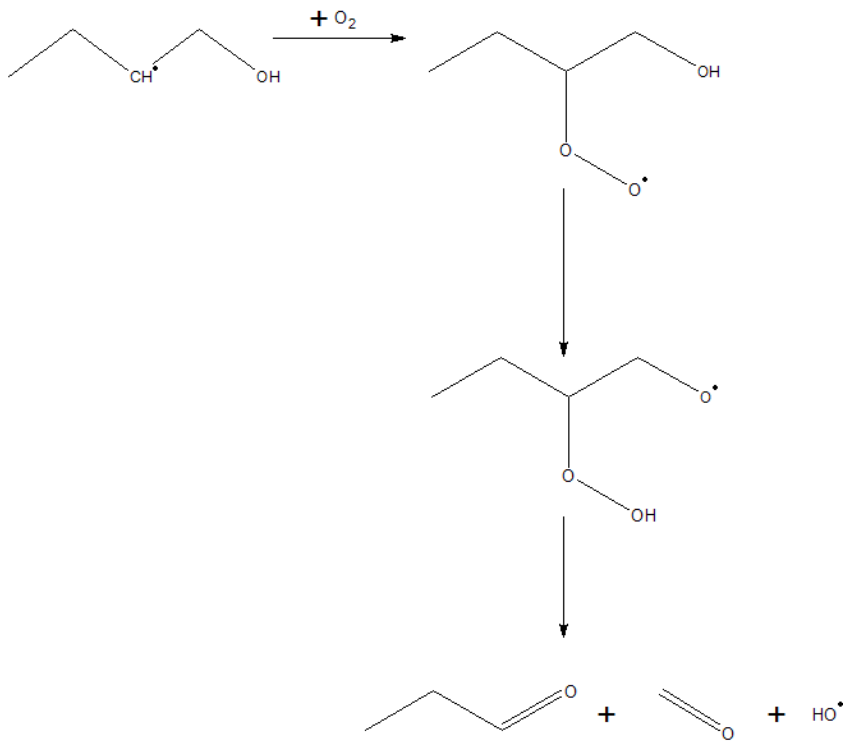

**Figure 4.20: Primary propanal formation pathway, from Ref. [43]**

In addition, Zhang et al.[43] showed that a significant amount of propanal formed prior to ignition. They suggested that the pathway responsible for the formation of propanal starts with molecular oxygen addition to the $\beta$-hydroxyalkyl radical of the fuel. Figure (4.20) shows the entire channel, reproduced from Ref [43]. Yang et al.[52] also showed propanal as a product of 1-butanol combustion, although they were not able to provide quantitative data. This channel is entirely absent from the NUI mechanism. Propanal subsequently decomposes into either formaldehyde + $C_2H_5$ or ethenal + $CH_3$, both of which are important radical channels. Thus, adding this channel may improve simulation results.

Figure 4.21: Initial fuel decomposition in the modified NUI mechanism at initial conditions of 800K, 15 atm, $\phi = 1.0$

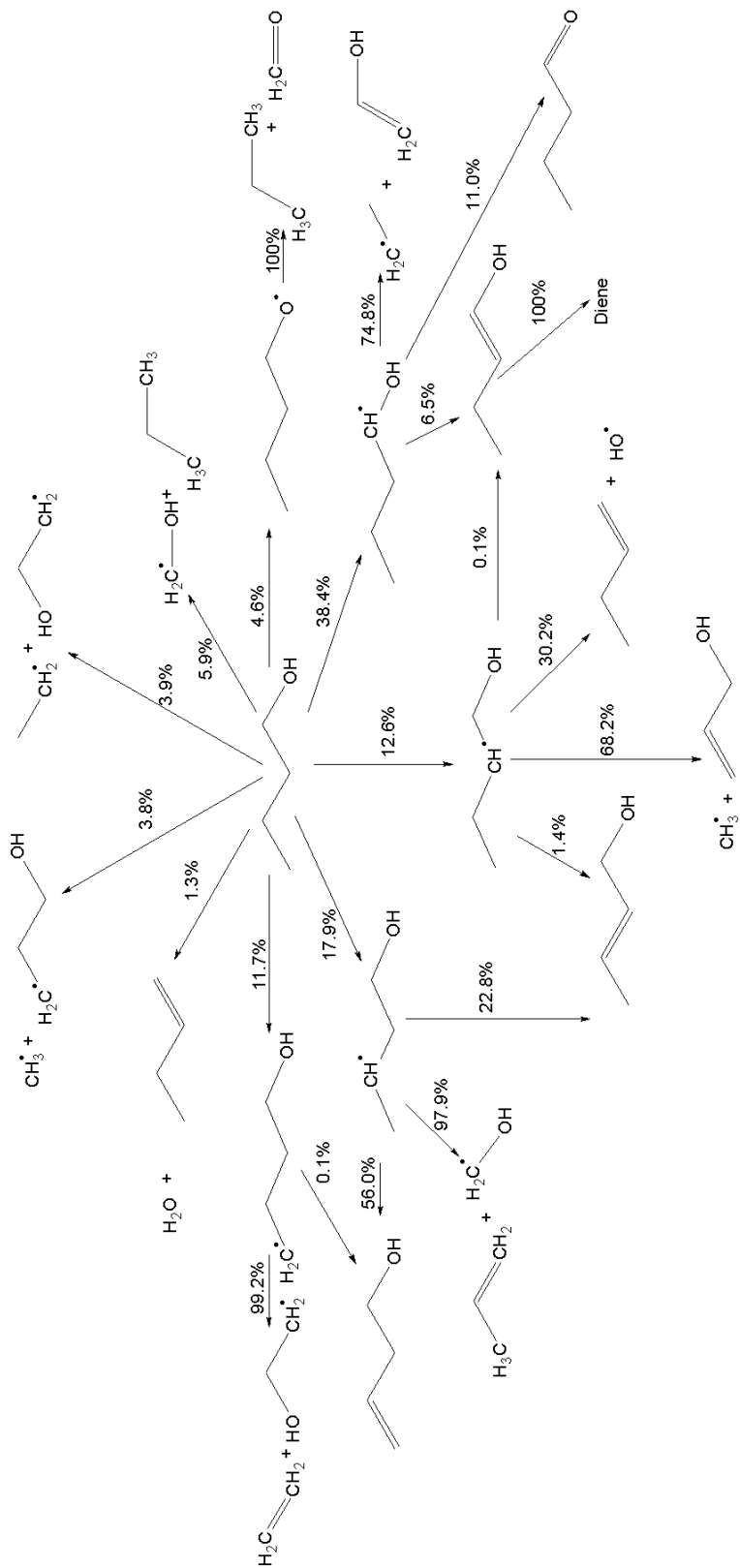

Figure 4.22: Initial fuel decomposition in the modified NUI mechanism at initial conditions of 1600K, 15 atm, $\phi = 1.0$

Figures (4.21 and 4.22) show results of the first formed species for the modified mechanism that has the rate of reaction 1353 multiplied by 1000. As expected in this case, the production of the $\alpha$-hydroxybutyl radical dominates, especially in the low temperature case. Comparing the low temperature case here and in Figure (4.16) shows marked similarity in the percent directed into the but-1-en-1-ol pathway. However, the butanal pathway has become much more important with the increasing concentration of $\alpha$-hydroxybutyl radical, at the expense of the unimolecular decomposition pathway.

Figure (4.13) shows that increasing the rate of R1353 does not further reduce the ignition delay, once the original rate for R1353 has been multiplied by about 500. This indicates that another reaction step has become the rate limiting step. Another brute force sensitivity analysis was conducted, this time on the modified mechanism (with the rate of R1353 multiplied by 1000), to help determine this new rate limiting step. The results are shown in Figure (4.23), with

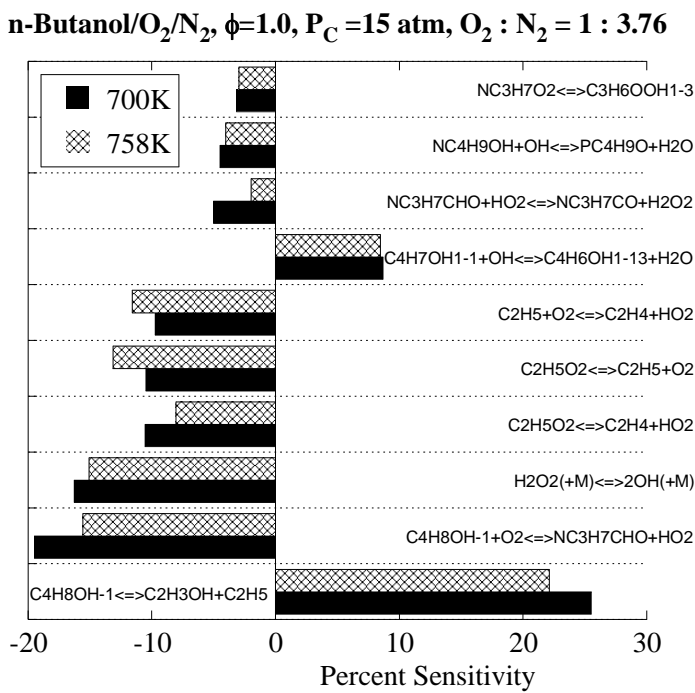

**Figure 4.23: Percent sensitivity of the modified NUI mechanism at initial conditions of Figure (4.13)**

the same initial conditions as in Figure (4.13). Note that the importance of R1353 has diminished significantly, as expected. It can be seen from Figure (4.23) that the system is quite sensitive to formation of butanal ($NC_3H_7CHO$) from the $\alpha$-hydroxybutyl radical. This reaction has a negative sensitivity, so increasing its rate decreases the ignition delay. The most sensitive reaction is the fission of the $\alpha$-hydroxybutyl radical. Increasing the rate of this reaction increases the ignition delay Also among the top 8 most sensitive reactions is a reaction which consumes butanal and one which consumes but-1-en-1-ol ($C_4H_7OH1$-1).The consumption of but-1-en-1-ol has a positive sensitivity, indicating that increasing the rate of this reaction will not reduce the ignition delay. This is most likely is due to the fact that the only two reactions of $C_4H_6OH1$-13 are back into but-1-en-1-ol and forming a diene, as shown in Figure 4.18.

Figure (4.24) shows the percent consumption of $\alpha$-hydroxybutyl by three major reactions as a function of multiplication factor of R1353 (R1353 is H-abstraction by $HO_2$ from the fuel to

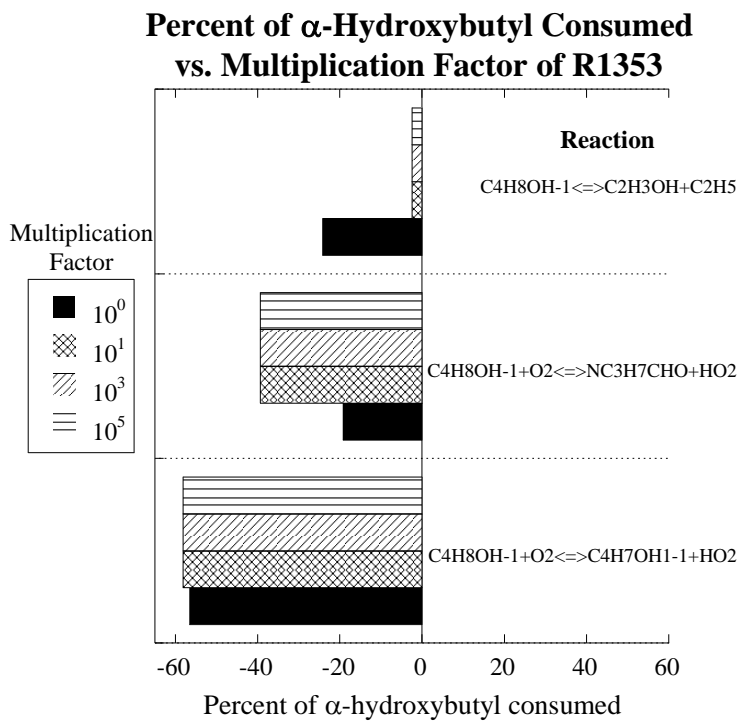

Figure 4.24: Percent consumption of $\alpha$-hydroxybutyl by three reactions at various multiplication factors of R1353. Initial conditions: 800K, 15 atm, $\phi$=1.0

form $\alpha$-hydroxybutyl). The initial conditions of the simulation were T = 800K, P = 15 atm and $\phi = 1.0$. All other consumption reactions for $\alpha$-hydroxybutyl were less than one percent. It is interesting that as soon as the rate of R1353 is increased, the percent consumed into the butanal pathway increased markedly and the pathway into $C_2H_3OH+C_2H_5$ decreased markedly, while the percent directed into but-1-en-1-ol remained approximately the same. It is also interesting that, even as the rate of R1353 is increased further, the percent of $\alpha$-hydroxybutyl consumed by each reaction remains approximately constant. This, together with the brute force analysis in Figure (4.23), suggests that the reaction into the butanal pathway is the rate limiting step.

All together, this analysis suggests that it will be necessary to modify more than just one reaction to bring simulations in line with experimental results. Determining these reactions can be accomplished by several means. Species profiles during the ignition process can help determine which pathways are most prominent in the fuel decomposition. This is the most straightforward way to determine which reaction rates need to be improved. Optical studies offer the benefit of real-time, in situ species measurements. This allows determining species profiles for more reactive species that may not survive long enough to be analyzed by methods such as GCMS. Finally, quantum calculations may provide some insight into the reaction rates relevant to the decomposition of butanol.

# Chapter 5.  Conclusions and Future Work

## 5.1  Conclusions

In this study, autoignition delays of *n*-butanol were measured at low to intermediate temperatures and at elevated pressures. In particular, temperature conditions from 650K – 900K were studied at pressures of 15 and 30 bar. Higher pressure experiments had shorter ignition delays than lower pressure cases.

In addition, independent variation of the concentration of fuel and oxygen at 3 equivalence ratios ($\phi = 0.5, 1.0,$ and $2.0$) has revealed the effect of each on the ignition delay. Increasing the fuel and oxygen concentration both decreased the ignition delay, while decreasing either increases the ignition delay.

Of particular note is the lack of NTC and two stage ignition delay in the temperature range studied. This result is of interest as all alkanes $C_3$ and greater show NTC in the temperature range studied here.

Simulated ignition delays computed using four mechanisms available in the literature are much longer than experimental ignition delays. In some cases, the discrepancy is several orders of magnitude. Although the reaction mechanisms had not been validated in the temperature and pressure range studied here, the degree of difference is still rather surprising.

To determine the source of the errors in the reaction mechanisms, the mechanism from the National University of Ireland at Galloway (NUI) was analyzed by two methods. The first, a brute force sensitivity analysis, shows that one reaction is much more important than the rest at low temperatures. This reaction is hydrogen abstraction from *n*-butanol by $HO_2$ to form the $\alpha$-hydroxybutyl radical. However, modifying the rate of just this reaction was insufficient to bring the simulations in line with the experiments.

Second, a reaction path analysis was performed to reveal the pathways by which the fuel decomposes. This analysis revealed several pathways which may need to have their rates adjusted and at least two pathways which were missing entirely. These pathways were propanal formation and tautomerization of but-1-en-1-ol to butanal.

## 5.2 Future Work

Thus, it seems very clear that more than one reaction will have to be modified to improve the results of simulations in the low temperature regime. Determination of which pathways or reactions to modify can be accomplished by several methods, including sampling of the reaction chamber and optical diagnostics for speciation measurements. To assist with comprehensive mechanism validation, speciation measurements can be performed on many experimental apparatuses, including jet-stirred reactors, rapid compression machines, spark-ignited engines and flame-based studies. Furthermore, quantum calculations can provide some guidance as to the values of rates for reactions.

In addition to improving currently available reaction mechanisms, one of the future goals of this project is to determine the suitability of *n*-butanol and its isomers as substitutes for petroleum-based transportation fuels. To this end, the isomers of butanol will be studied in the RCM to characterize their ignition delays. Finally, since butanol is likely to be blended with gasoline during a transitionary phase, blends of butanol with other fuels will be studied.